\documentstyle[12pt,epsfig,cite]{article} 
 
\def\slashchar#1{\setbox0=\hbox{$#1$}\dimen0=\wd0%
\setbox1=\hbox{/}\dimen1=\wd1%
\ifdim\dimen0>\dimen1%
\rlap{\hbox to 
\dimen0{\hfil/\hfil}}#1\else                                        
\rlap{\hbox to \dimen1{\hfil$#1$\hfil}}/\fi} 

\begin{document} 
\textwidth 17.0cm 
\textheight 22.0cm 
 
\newcommand{\be}{\begin{equation}} 
\newcommand{\ee}{\end{equation}} 
\newcommand{\ba}{\begin{eqnarray}} 
\newcommand{\ea}{\end{eqnarray}} 
\def\eqn#1{(\ref{#1})} 
\newcommand{\cO}{{\cal O}} 
\newcommand{\cL}{{\cal L}} 
\newcommand{\cQ}{{\cal Q}} 
\newcommand{\cA}{{\cal A}} 
 
%
%
 
\begin{titlepage} 
\begin{flushright} {FTUV/00-0717 \\ IFIC/00-30 \\ }
\end{flushright} 
\vspace{0.8cm} 
\begin{center} {\Large\bf Final State Interactions in Kaon Decays   
}\\ 
\vspace{2.0cm} {\bf Elisabetta Pallante${}^a$\footnote{e--mail: 
     pallante@ecm.ub.es} and  Antonio Pich${}^b$\footnote{e--mail: 
Antonio.Pich@uv.es}     } 
\\[0.5cm] 
 { ${}^a$ Facultat de F\'{\i}sica, Universitat de Barcelona, Diagonal 
647,\\ E-08028 Barcelona, Spain  } 
\\[0.3cm] 
 { ${}^b$ Departament de F\'{\i}sica Te\`orica, IFIC,  Universitat de 
Val\`encia -- CSIC, \\  Apt. Correus 22085, E-46071 Val\`encia, Spain 
}\\[0.5cm] 
\end{center} 
\vfill 
\begin{abstract} 
We quantify the important effect of strong final 
state interactions in the  weak $K\to 2\pi$ amplitudes, 
using the measured $\pi$-$\pi$ phase shifts with $J=0$ and $I=0,2$. 
The main results of this analysis, with their implications  for 
$\varepsilon'/\varepsilon$ and the $\Delta I =1/2$ rule, 
have been already presented in a previous paper \cite{SHORT}. 
Here we provide a detailed formal derivation of those results 
and further discuss 
the Standard Model prediction of $\varepsilon'/\varepsilon$. 
\end{abstract} 
\vspace{1cm} 
\hspace{0.6cm}{\small PACS numbers: 13.25.Es, 11.30.Er, 13.85.Fb, 
11.55.Fv } 
\vfill 
 
\end{titlepage} 
\setcounter{equation}{0} 
\setcounter{figure}{0} 
\renewcommand{\theequation}{\arabic{section}.\arabic{equation}} 
\setcounter{equation}{0} 
 
 
\section{Introduction} 
 
It is well known that, at centre--of--mass energies around the kaon 
mass, the strong S--wave $\pi$--$\pi$ scattering generates a large 
phase-shift difference 
$\left(\delta^0_0 - \delta^2_0\right)(M_K^2)= 45^\circ\pm 6^\circ$ 
between the $I=0$ and $I=2$ partial waves \cite{GM}.  In the usual 
description of $K\to 2\pi$ decays, this effect is explicitly taken 
into account, through the following decomposition of the relevant 
isospin amplitudes with $I=0$ and $I=2$: 
\be\label{eq:AIdef} 
{\cal A}_I \,\equiv\, A\left[ K\to (\pi\pi )_I\right] 
\,\equiv\, A_I \; e^{i\delta^I_0} \; . 
\ee 
It is also known 
\cite{BBG_1,TR88,BE:89,IS90,BDJS:90,NS:91,LMZ:97,KNS:99,NI:99,KA91} 
that final state 
interactions (FSI) play an important role in the observed enhancement 
of the $I=0$ decay amplitude, $A_0/A_2\approx 22.2$. 
The presence of  such a large phase-shift difference clearly signals a 
corresponding dispersive FSI effect in the moduli of the isospin amplitudes, 
because the real and imaginary parts are related by analyticity and unitarity. 
 
The size of the induced FSI correction can be roughly estimated from the 
available one-loop analyses of $K\to 2\pi$ \cite{KA91,BPP,JHEP} in 
Chiral Perturbation Theory ($\chi$PT). 
At lowest order in the momentum expansion, $O(p^2)$, 
the decay amplitudes do not contain any strong phase. 
Those phases originate in the 
final rescattering of the two pions and, therefore, are generated by 
chiral loops which are of higher order in momenta. 
Since the strong phases are quite large, one should expect large 
higher--order unitarity corrections. 
The one-loop calculations \cite{KA91,BPP,JHEP} 
show in fact that the pion 
loop diagrams provide an important enhancement of the $A_0$ amplitude, 
of about 40\%. 
However, the phase-shift $\delta^0_0$ predicted  by the one-loop 
calculation is still lower than its measured value, which indicates that 
a further enhancement could be expected at higher orders. 
 
Although the importance of FSI in $K\to 2\pi$ has been known for more than 
a decade, their impact on the direct CP-violating parameter 
$\varepsilon'/\varepsilon$ has been overlooked in the 
so--called {\it Standard Model predictions} of this parameter, presented 
in refs. \cite{Buras} and \cite{Lattice}. 
Not surprisingly, those predictions fail to reproduce the experimental 
measurements \cite{exp}. 
 
The lattice investigations of kaon decay amplitudes 
have been only able, up to now, to compute the one-pion 
$\langle\pi|{\cal H}_{\Delta S=1}|K\rangle$ matrix elements. 
In order to get the physical two-pion decay amplitudes, they 
rely on the lowest--order $\chi$PT relation between $K\to\pi$ and 
$K\to 2\pi$, 
which, as mentioned before, does not include any FSI and 
underestimates the $I=0$ \ $K\to 2\pi$ amplitude by at least 40\%. 
 
In refs. \cite{Buras}, the large--$N_C$ limit is used to fix the 
CP--violating $K\to 2\pi$ decay amplitudes. Since the 
strong phases $\delta^I_J$ are zero at leading order in the $1/N_C$ 
expansion, the FSI enhancement has not been taken into account, either. 
Other approaches \cite{Trieste,Dortmund} 
include some one-loop corrections and find larger values for the 
$A_0$ amplitude. Although those are model--dependent estimates, 
they provide an indication of the importance of higher--order pion--loop 
contributions. 
 
A proper way to account for the FSI effects has been addressed in 
ref.~\cite{SHORT}, where it has been shown that the strong rescattering 
of the two final pions generates a large enhancement of 
$\varepsilon'/\varepsilon$. The resulting Standard Model prediction 
\cite{SHORT,PPS:00}, 
\be\label{eq:epsth} 
\left.\mbox{Re}\left(\varepsilon'/\varepsilon\right)\right|_{SM} = 
(17\pm 6)\times 10^{-4} \, , 
\ee 
is in good agreement with the present experimental world average 
\cite{Ceccuchi} 
\be\label{eq:exp} 
\left.\mbox{Re}\left(\varepsilon'/\varepsilon\right)\right|_{exp} = 
(19.3\pm 2.4)\times 10^{-4} \, . 
\ee 

In the following we provide a detailed discussion of the approach 
advocated in ref.~\cite{SHORT} and further study its implications 
for $\varepsilon'/\varepsilon$ and the $\Delta I=1/2$ rule. 
The paper is organized as follows. In section \ref{FSI} we 
formulate the Omn\`es problem for a general amplitude with two 
pions in the final state and derive 
its solution (for any number of subtractions). It is shown how the Omn\`es 
dispersive factor, solution of the Omn\`es problem, provides 
an all--order resummation 
of the infrared chiral logarithms that contribute to FSI. 
This is a universal process--independent factor, 
which only depends on the quantum numbers 
($I$ and $J$) of the final two--pion state. 
There is of course a polynomial (local) ambiguity, which encodes the 
process--dependent ultraviolet dynamics. 
 
To clarify the physics involved in the Omn\`es resummation, we 
present in section~\ref{Scalar} explicit results for the 
scalar pion form factor. This quantity is known to two loops 
in the chiral expansion and, therefore, provides a simple example 
where the power of our approach can be easily shown and the 
uncertainties quantified. 
The much more involved case of $K\to 2\pi$ transitions is discussed 
in section~\ref{K2pi}, while section~\ref{eps} presents 
the Standard Model prediction of $\varepsilon'/\varepsilon$. 
We conclude in section \ref{CONC} with a few summarizing comments. 
We have relegated to the appendices the details on experimental 
$\pi$--$\pi$ phase shifts, 
one-loop $\chi$PT results for $K\to 2\pi$ 
and some remarks on recent literature on the subject.

\section{Omn\`es approach to FSI} 
\label{FSI}\setcounter{equation}{0} 
 
Let us consider a generic amplitude (or form factor) $A^I_J(s)$, 
with two pions in the final state 
which have total angular momentum and isospin given by $J$ and $I$, 
respectively, and invariant mass 
$s\equiv q^2\equiv \left(p_1 + p_2\right)^2$. 
As indicated by the name, FSI refer to the final dynamics of the two pions 
and not to the particular process leading to this final state. 
Therefore, we will not specify the physical amplitude and will look for a 
way to resum the strong $\pi$--$\pi$ interactions to all orders in the 
chiral expansion.

Let us define the amplitude $A^I_J(s)$ being analytic on the complex $s$ 
plane except for a cut $L\equiv [4 M_\pi^2, \infty )$ 
along the real positive $s$ axis. 
For real values $s<4 M_\pi^2$ the amplitude is real; 
this implies that the values of the amplitude above and below the cut 
are complex conjugate of each other: 
$A^I_J(s+i\epsilon )  = A^I_J(s-i\epsilon )^\ast$. 
 Above the threshold, $s\geq 4 M_\pi^2$, 
 $A^I_J(s)$ has a discontinuity across the cut and develops an absorptive 
(imaginary) part. 
 
Cauchy's theorem implies that $A^I_J(s)$ can be written 
as a dispersive integral along the physical cut: 
\be\label{eq:DISP} 
A^I_J(s) = {1\over \pi}\int_{L}~ d z \; 
{{\mbox{Im}}\, A^I_J(s)\over z-s-i\epsilon} 
\, +\, {\mbox{subtractions}}\, . 
\ee 
The convergence of the dispersive integral is dictated by the specific 
form of the function $A^I_J(s)$. 
Depending on the particular asymptotic behaviour of $A^I_J(s)$ 
at the extremes of the cut $L$, 
a number of subtractions has to be performed to make the integral convergent. 
 
Let us further assume that $A^I_J(s)$ corresponds to some weak or 
electromagnetic transition, in the presence of strong interactions. 
Thus, above the cut, 
$A^I_J(s+i\epsilon ) = \langle (\pi\pi)^I_J\vert   O\vert i(q)\rangle $ 
where $O$ could be some effective (low--energy) electroweak Hamiltonian 
or a current. 
Working to first order in the small electroweak coupling, 
the unitarity condition allows then to write the 
imaginary part of $A^I_J(s)$ as a sum over the contributions from all 
possible on--shell intermediate states which couple 
to the initial and the final state (properly normalized in momentum space): 
\be\label{eq:ASIM} 
\mbox{Im}\, A^I_J(s+i\epsilon )\, =\, {1\over 2} 
 \sum_n \,\langle  (\pi\pi)^I_J\vert T^\dagger \vert n \rangle\; 
\langle  n \vert O \vert i(q) \rangle \, , 
\ee 
where $T$ is the scattering $T$--operator. 
To derive this result, one makes use of Time--Reversal invariance; 
we will comment later on the proper way to bypass this assumption 
when analyzing CP--violating observables. 
 
Note that since $\mbox{Im}\, A^I_J$ is real, 
also the r.h.s. of eq.~(\ref{eq:ASIM}) is real. 
Below the first inelastic threshold, only the elastic channel contributes 
to the sum; one gets then: 
\ba\label{eq:WASIM_EL} 
\mbox{Im}\, A^I_J \, =\, \left(\mbox{Im}\, A^I_J\right)_{2\pi} &=& 
 e^{-i\delta^I_J} \,\sin{\delta^I_J}\; A^I_J  \, =\, 
e^{i\delta^I_J} \,\sin{\delta^I_J}\; A^{I\ast}_J  
\nonumber\\ &=& 
\sin{\delta^I_J}\; \vert A^I_J\vert \, =\, \tan{\delta^I_J}\; 
\mbox{Re}\, A^I_J\, , 
\ea 
implying that the phase of the decay amplitude 
$A^I_J(s+i\epsilon ) = \vert A^I_J(s)\vert \exp{\left(i\delta^I_J\right)}$ 
is equal to  the phase of $T^I_J$, the $\pi\pi\to\pi\pi$ partial--wave  
scattering amplitude  (Watson's theorem \cite{Watson}). 
Eq.~(\ref{eq:WASIM_EL}) expresses the imaginary part of the amplitude 
$A^I_J(s)$ in terms of the 
amplitude itself, or its real part, and the $\pi\pi$ phase--shift 
$\delta^I_J(s)$. 
 
Inserting eq.~(\ref{eq:WASIM_EL}) in the dispersion relation 
(\ref{eq:DISP}), one obtains an integral equation for $A(s)$ of 
the Omn\`es type, 
\be\label{eq:disp} 
A^I_J(s)  = \sum_{k=0}^{n-1} {(s-s_0)^k\over k!} 
\left. {d^k A^I_J\over ds^k}\right|_{s=s_0} 
+ \frac{(s-s_0)^n}{\pi} \int^{\infty}_{4M^2_\pi} \, \frac{dz}{(z-s_0)^n} 
\, {\tan{\delta^I_J(z)} \, \hbox{Re}A^I_J(z)\over z-s-i\epsilon} \, , 
\ee 
which has the well--known Omn\`es \cite{TR88,MU:53,OM:58,PLBFF} solution: 
\be\label{eq:NSUB} 
A^I_J(s) \, = \, Q^I_{J,n}(s,s_0)~\exp{\left\{I^I_{J,n}(s,s_0)\right\}} 
\, , 
\ee 
where 
\be\label{eq:In} 
I^I_{J,n}(s,s_0) \,\equiv\, {(s-s_0)^{n} \over \pi} 
\int^\infty_{4M^2_\pi}~ 
{dz\over (z-s_0)^{n}}  \; 
{\delta^I_J(z) \over z-s-i\epsilon} \, , 
\ee 
$Q^I_{J,0}(s,s_0)\equiv 1$ \ 
and 
\be\label{eq:NSUB_Q} 
Q^I_{J,n}(s,s_0) \,\equiv\, \exp{\left\{\sum_{k=0}^{n-1}\, 
{ (s-s_0)^{k} \over k!}\; {d^k \over d s^k}   
\left.\log{\left\{A^I_J(s)\right\}} \right|_{s=s_0}\right\}} \, , 
\qquad (n\geq 1)\, . 
\ee 

Strictly speaking, this equation is only valid below the first inelastic 
threshold  ($s \leq 16 M_\pi^2)$. However, the contributions from 
higher--mass intermediate states are suppressed  by phase space. 
The production of a larger number of meson pairs is also of 
higher order in the chiral expansion. 
 
We have written the most general result, for a given number of 
subtractions $n$, performed at a generic subtraction point $s_0$ 
outside the physical cut. 
The dispersive integral $I^I_{J,n}(s,s_0)$ 
is uniquely determined up to a polynomial ambiguity 
(that does not produce any imaginary part of the amplitude), 
which depends on the number of subtractions and the subtraction point. 
This can be readily seen through 
the use of the following iterative formula for the real 
part of $I^I_{J,n}(s,s_0)$: 
\be\label{eq:ITER} 
\mbox{Re}\,  I^I_{J,n}(s,s_0) \, =\, 
\mbox{Re}\,  I^I_{J,n-1} (s,s_0) \, - \, 
(s-s_0)^{n-1}\lim_{s\to s_0} {\mbox{Re}\, I^I_{J,n-1}(s,s_0) 
\over (s-s_0)^{n-1}} 
\, , 
\ee 
where the second term on the r.h.s. depends on $s$ only through the 
polynomial factor $(s-s_0)^{n-1}$. 
The non--polynomial part of  $I^I_{J,n}(s,s_0)$, 
containing the infrared chiral logarithms, does not have any 
dependence on the number of subtractions or the subtraction point. 
The polynomial ambiguity is of course canceled by the 
subtraction function  $Q^I_{J,n}(s,s_0)$.

Thus, the Omn\`es solution predicts the chiral logarithmic corrections 
in a universal way, independently of the number of subtractions or the 
subtraction point, and 
provides their exponentiation to all orders in the chiral expansion. 
The polynomial ambiguity of $I^I_{J,n}(s,s_0)$ and the subtraction 
function  $Q^I_{J,n}(s,s_0)$ can be fixed, at a given order in the 
chiral expansion, by matching the Omn\`es formula (\ref{eq:NSUB}) 
with the $\chi$PT prediction of $A^I_J(s)$. 
It remains a polynomial ambiguity at higher orders. 
 
A special case, which turns out to be relevant in the treatment of 
the weak $K\to 2\pi$ amplitudes, is the one where the amplitude 
$A^I_J(s)$ has a zero of a given order $p$ at some point $s=\zeta$. 
In this case, once the zero is factorized 
through the relation $A^I_J(s) = (s-\zeta)^p \,\overline{A^I_J}(s)$, 
the dispersive 
representation of eq.~(\ref{eq:NSUB}) is valid for the function 
$\overline{A^I_J}(s)$, so that 
\be\label{eq:NSUB_ZERO} 
A^I_J(s) \, =\, (s-\zeta)^p \;\overline{A^I_J}(s) \, =\, 
(s-\zeta)^p \; \overline{Q}^I_{J,n}(s,s_0)\; 
\exp{\left\{ I^I_{J,n}(s,s_0)\right\}}\, , 
\ee 
with $\overline{Q}^I_{J,n}(s,s_0)$ the analogous of the expansion 
 $Q^I_{J,n}(s,s_0)$ of eq.~(\ref{eq:NSUB_Q}) for the function 
 $\overline{A^I_J}(s)$ and 
$I^I_{J,n}(s,s_0)$ as defined in  (\ref{eq:In}).

\section{The scalar pion form factor} 
\label{Scalar}\setcounter{equation}{0}

The scalar form factor of the pion is the simplest quantity where the 
Omn\`es problem can be solved \cite{DGL:90,GM:91} 
in order to resum final state interactions 
of a two-pion state with total angular momentum $J=0$ and isospin $I=0$. 
It is defined by the matrix element of the $SU(2)$ quark scalar density 
\be 
\langle\pi^i(p^\prime )\vert \bar{u}u+\bar{d}d\vert\pi^k(p)\rangle 
\,\equiv\, \delta^{ik}\, F_S^{\pi}(t)\, . 
\label{eq:SFF} 
\ee 

At low momentum transfer, $\chi$PT provides a systematic expansion 
of $F_S^{\pi}(t)$ in powers of $t\equiv (p^\prime - p)^2$ 
and the light quark masses \cite{GL:84,GL:85}: 
\be 
F_S^{\pi}(t) = F_S^{\pi}(0)\left\{ 1+ g(t) +O(p^4)\right\}\, . 
\label{eq:SFF_2} 
\ee 
The value at $t=0$ coincides with the pion sigma term. It can 
be written as an expansion in powers of the light quark masses as 
follows: 
\be 
F_S^{\pi}(0) = \left( 
  {\partial\over\partial m_u} + {\partial\over\partial m_d} 
  \right ) M_\pi^2 = 2 B_0 + O(m_q) \, ,  
\ee 
where  
$B_0$ is a coupling of the lowest--order $\chi$PT Lagrangian which is related 
to the quark--antiquark vacuum condensate. 
The $O(p^2)$ correction $g(t)$ contains contributions 
from one-loop diagrams and tree--level terms of the $O(p^4)$ $\chi$PT 
Lagrangian. It is given by \cite{GL:84,GL:85}: 
\ba\label{eq:g_scalar} 
g(t) &=& {t\over f^2}\left\{ \left (1-{M_\pi^2\over 2t}\right ) 
\bar{J}_{\pi\pi}(t) + {1\over 4}\,\bar{J}_{KK}(t) + {M_\pi^2\over 18 t} 
\,\bar{J}_{\eta\eta}(t)  +4(L_5^r+2L_4^r)(\mu ) \right .\nonumber\\ 
&&\left.\quad\mbox{} + {5\over 4(4\pi )^2} 
\left (\ln{\mu^2\over M_\pi^2} -1\right ) -  {1\over 4(4\pi )^2} 
\ln{M_K^2\over M_\pi^2} \right\}\, , 
\ea 
with $f\approx f_\pi = 92.4$ MeV. 
 
Only two terms of the strong $\chi$PT Lagrangian of 
$O(p^4)$ contribute to $g(t)$. The $\mu$ dependence of their corresponding 
chiral couplings $L_4^r(\mu )$ and $L_5^r(\mu )$ exactly cancels the 
one from the chiral loops appearing through the logarithm 
$\ln{(\mu^2/ M_\pi^2)}$. 
At the standard reference value $\mu = M_\rho$, 
$[L_5^r +2L_4^r](M_\rho) = (0.8\pm 1.1)\times 10^{-3}$ 
\cite{EC:95,PI:95,ME:93}.
 
The functions $\bar{J}_{\pi\pi}(t)$, $\bar{J}_{KK}(t)$ and 
$\bar{J}_{\eta\eta}(t)$ 
are ultraviolet finite and, together with the logarithms, they are produced 
by the one-loop exchange 
of $\pi\pi$, $K\bar K$ and $\eta\eta$ intermediate states. 
They have the form: 
\be 
\bar{J}_{PP}(t) = {1\over (4\pi )^2}\left\{ 2 - \sigma_P\ln{\left ( 
{\sigma_P +1\over \sigma_P -1}\right )}\right\} 
\qquad ; \qquad 
\sigma_P\equiv\sqrt{1-{4 M_P^2\over t}}\, . 
\label{eq:Jb_PP} 
\ee 

Below the first inelastic threshold, 
the absorptive part of the scalar form factor is generated by 
$\pi\pi$ exchange, through the one-loop function $\bar{J}_{\pi\pi}(t)$. 
Thus, all non-analytic contributions originate from the final 
rescattering of the two pions and could be studied within the chiral 
$SU(2)\otimes SU(2)$ framework \cite{GL:84}, 
where the $K$ and $\eta$ modes are integrated out. 
In fact, for values of $t$ such that $t\ll 4 M_P^2$, 
\be 
\bar{J}_{PP}(t) = {1\over (4\pi )^2}\left [ {1\over 6} {t\over M_P^2} 
+ {1\over 60} {t^2\over M_P^4} + \ldots \right ]\, , 
\ee 
implying that, below the $P\bar P$ threshold, 
$\bar{J}_{PP}(t)$ has a very smooth behaviour and is strongly suppressed. 
In the case of the pion scalar form factor this means that at values of 
$t\ll 4 M_K^2\, , 4 M_\eta^2$ the 
one-loop functions $\bar{J}_{KK}(t)$ and $\bar{J}_{\eta\eta}(t)$ 
only give small analytic corrections, which are 
numerically negligible with respect to the local 
$(L_5^r+2L_4^r)(M_\rho )$ contribution. 
 
The two loop corrections to the pion form factor 
have been already computed \cite{BGT:98}. However, we 
prefer to keep the discussion at the one-loop level only, in order to 
make easier the comparison with $K\to \pi\pi$ where two-loop 
corrections are not yet available. 
 
Let us consider now the  Omn\`es problem for $F_S^\pi (t)$ 
and fix the subtraction polynomial performing a matching 
with the one-loop $\chi$PT result. 
We can write the Omn\`es solution in the form: 
\be 
F_S^{\pi}(t)\, =\, \Omega_0 (t,t_0) \cdot F_S^{\pi}(t_0) 
\,\approx\, 
\Omega_0 (t,t_0)\cdot F_S^{\pi}(0)\,\left\{ 1+ g(t_0)\right\} \, . 
\label{eq:OMNES_SFF} 
\ee 
Thus, knowing the form factor at some low--energy subtraction point 
$t_0$, where the momentum expansion can be trusted, 
the Omn\`es factor $\Omega_0 (t,t_0)$ provides an evolution of 
the result to higher values of $t$, through the exponentiation 
of infrared effects related to FSI. 
The once--subtracted solution reads: 
\be 
\Omega_0^{(1)} (t,t_0) 
\, =\, \exp\left\{ {t-t_0\over \pi}\int_{4M_\pi^2}^{\bar{z}}\, 
{dz\over z-t_0}\, {\delta_0^0(z)\over z-t-i\epsilon} \right\} 
\,\equiv\, 
\Re_0^{(1)} (t,t_0) \; e^{i\delta^0_0(t)}\, . 
\label{eq:OMNES_F} 
\ee 
We have split the integral into its real and 
imaginary part, making  explicit that the phase of the Omn\`es factor 
is just the original phase-shift $\delta^0_0(t)$. The Omn\`es exponential 
generates its corresponding dispersive factor $\Re_0^{(1)} (t,t_0)$.

The integral has been cut at the 
upper edge $\bar{z}$, which represents the first inelastic threshold. 
Above $\bar{z}$ the representation (\ref{eq:OMNES_F}) 
is no longer valid and a coupled--channel analysis is required  
to solve the  Omn\`es problem. From the behaviour of  
the S--wave $\pi\pi$ 
phase-shift $\delta_0^0(z)$ (see appendix~\ref{App:phaseshift}), 
it is immediate to 
infer that the elastic integral evaluated 
up to $\bar{z}\sim 1$ GeV${}^2$ 
will slightly underestimate the exact result obtained with the 
inclusion of inelastic contributions. We shall discuss this point in 
more detail later. 
 
The solution for $F_S^{\pi}(t)$ 
given in eq. (\ref{eq:OMNES_SFF}), being a physical quantity, 
 must be independent of the 
subtraction point $t_0$, while the  Omn\`es factor and the 
amplitude in front of it do depend on $t_0$. 
For illustrative purposes, we take $t=M_K^2$ 
(the scale relevant for $K\to\pi\pi$) and 
show in Table~\ref{table:SFF} the resulting value of 
$|F_S^{\pi}(M_K^2)/F_S^{\pi}(0)|$ for different subtraction points 
$t_0 = 0,\, M_\pi^2,\, 2 M_\pi^2, \, 3 M_\pi^2, \, 4M_\pi^2$ and $M_K^2$. 
The upper limit of the integration range has been fixed at 
$\bar{z} = 1$ GeV${}^2$. 
Our Omn\`es integral (\ref{eq:OMNES_F}) 
is not defined for $t_0 =M_K^2$, because it lies above the threshold 
of the non--analyticity cut; however, 
by its definition in eq.~(\ref{eq:OMNES_SFF}), 
$\Omega (M_K^2,t_0=M_K^2) = 1$ holds. 
 
\begin{table}[htb] 
\begin{center} 
\begin{tabular}{|c|c|c|c||c|c|} 
\hline 
$t_0$  & $g(t_0)$ & $\Re_0^{(1)} (M_K^2,t_0)$ & 
$\Re_0^{(2)} (M_K^2,t_0)$ & 
\multicolumn{2}{c|}{$|F_S^{\pi}(M_K^2)/F_S^{\pi}(0)|$} \\ 
 \cline{5-6} &&  && 
$n=1$ & $n=2$ \\ 
\hline 
 0           &  0    & (1.23) \ \ 1.45 & (1.55) \ \ 1.56 & 
  (1.23) \ \ 1.45 & (1.55) \ \ 1.56 \\ 
$M_\pi^2$    & 0.042 & (1.21) \ \ 1.40 & (1.47) \ \ 1.44 & 
  (1.26) \ \ 1.46 & (1.53) \ \ 1.50 \\ 
$2 M_\pi^2$  & 0.091 & (1.17) \ \ 1.34 & (1.38) \ \ 1.31 & 
  (1.28) \ \ 1.46 & (1.51) \ \ 1.43 \\ 
$3 M_\pi^2$  & 0.15  & (1.12) \ \ 1.26 & (1.26) \ \ 1.13 & 
  (1.29) \ \ 1.45 & (1.45) \ \ 1.30 \\ 
$4 M_\pi^2$  & 0.26  & (1.03) \ \ 1.11 & ---  & 
  (1.30) \ \ 1.40 &  --- \\ 
$M_K^2$  & $0.54 - 0.46\, i$ & $\equiv 1$ & 
      $\equiv 1$ &  1.61 & 1.61  \\ 
\hline 
\end{tabular} 
\end{center} 
\protect\caption{ 
The one-loop function $g(t_0)$, 
the Omn\`es factor $\Re_0^{(n)}(t,t_0)$ and the modulus of 
$F_S^{\pi}(t)/F_S^{\pi}(t_0)$ 
are shown at $t=M_K^2$ for different values of the 
subtraction point $t_0\in [0,M_K^2]$ 
and for $n=1,\, 2$ subtractions. 
At a given $t_0$, the first value 
(within brackets) is obtained with 
the $O(p^2)$ $\chi$PT prediction for $\delta_0^0$, while the 
fit to the experimental phase-shift data has been used in the second one. 
The integrals have been cut at $\bar{z} = 1$ GeV${}^2$.} 
\label{table:SFF} 
\end{table} 
 
The dominant contributions to $g(t_0)$ 
come from the logarithms and the $\pi\pi$ one-loop function 
$\bar{J}_{\pi\pi}(t)$. 
The corrections from $\bar{J}_{KK}(t_0)$ and $\bar{J}_{\eta\eta}(t_0)$ 
stay within $1\%$ of the total one--loop correction at all non-zero 
subtraction points, while the local $[L_5^r +2L_4^r](M_\rho)$ 
term gives a contribution smaller than 10\%. 
The $\chi$PT calculation of $F_S^\pi(t_0)$ is obviously better 
at lower values of $t_0$ where the one-loop correction $g(t_0)$ 
is smaller. At $t_0 = 4 M_\pi^2$ a sizeable 26\% effect is 
already found, while at $t_0 = M_K^2$ the correction is so large 
than one should worry about higher--order contributions. 
The Omn\`es exponential allows us to predict $F_S^\pi(M_K^2)$ in a 
much more reliable way, through the evolution of safer results 
at lower $t_0$ values.

The resulting values for the scalar form factor remain very stable 
within the whole range of subtraction points. 
Increasing the value of $t_0$  one just moves 
higher--order $\chi$PT corrections from the Omn\`es factor to the 
amplitude $F_S^\pi(t_0)$. 
At $t_0 = M_\pi^2 $ the one-loop 
corrections are still almost zero, while the 
Omn\`es factor contains all the higher--order effects. 
At the highest possible subtraction point $t_0 =4 M_\pi^2 $, i.e. the 
threshold of the non--analyticity cut, the bulk of the higher--order 
corrections has been moved to $F_S^\pi(t_0)$, 
while the Omn\`es factor approaches one, the 
value that it assumes at $t_0 = M_K^2$, by construction. 
 
Together with the more accurate results obtained with the 
experimental phase-shifts, we have shown in Table~\ref{table:SFF} 
(within brackets) 
the corresponding numerical values using the lowest--order 
$\chi$PT prediction for $\delta^0_0(z)$ in eq.~(\ref{eq:CHPT_delta}). 
As expected, the $O(p^2)$ $\chi$PT approximation to $\delta^0_0(z)$ 
underestimates the dispersive integral and, therefore, the 
true results, which are obtained with the experimental phase-shifts. 
 
As shown in Table~\ref{table:SFF_Z}, the dispersive correction factors 
$\Re_0^{(1)}(M_K^2,t_0)$ increase with increasing values of the upper 
integration limit $\bar{z}$. This 
clearly indicates that we are underestimating the FSI effect. However, 
one cannot trust the numerical results obtained for 
$\bar{z} > 1$ GeV${}^2$ 
because a coupled--channel analysis is required 
above the inelastic threshold.

\begin{table}[htb] 
\begin{center} 
\begin{tabular}{|c||c|c|c||c|c|c|} 
\hline 
& \multicolumn{3}{c||}{$\Re_0^{(1)} (M_K^2,t_0)$}  & 
\multicolumn{3}{c|}{$\Re_0^{(2)} (M_K^2,t_0)$} 
\\ 
\cline{2-7} & 
$\bar{z}=1$  & $\bar{z}=2$ & $\bar{z}=3$ & 
$\bar{z}=1$  & $\bar{z}=2$ & $\bar{z}=3$ 
\\ 
\hline 
$t_0=0$          & 1.45 & 1.58 & 1.62 & 1.56 & 1.58 & 1.59 
\\ 
$t_0=M_\pi^2$    & 1.40 & 1.51 & 1.55 & 1.44 & 1.46 & 1.47 
\\ 
$t_0= 2 M_\pi^2$ & 1.34 & 1.44 & 1.47 & 1.31 & 1.32 & 1.33 
\\ 
$t_0= 3 M_\pi^2$ & 1.26 & 1.35 & 1.38 & 1.13 & 1.14 & 1.14 
\\ 
$t_0= 4 M_\pi^2$ & 1.11 & 1.19 & 1.21 &  --- &  --- & --- 
\\ 
\hline 
\end{tabular} 
\end{center} 
\protect\caption{Dependence of $\Re_0^{(n)} (M_K^2,t_0)$ 
on the upper edge $\bar{z}$ (in GeV${}^2$ units) 
of the dispersive integral, 
for various choices of $t_0$ and $n=1,\, 2$. 
The fit to the experimental data for $\delta_0^0$ 
has been used.} 
\label{table:SFF_Z} 
\end{table} 
 
We can suppress the sensitivity to the high integration range by using 
a twice--subtracted dispersion relation. The corresponding 
Omn\`es factor is given by: 
\ba 
\Omega_0^{(2)} (t,t_0)  & = & 
\exp\left\{ (t-t_0) {g'(t_0)\over 1 + g(t_0)} + 
{(t-t_0)^2\over \pi}\int_{4M_\pi^2}^{\bar{z}}\, 
{dz\over (z-t_0)^2}\, {\delta_0^0(z)\over z-t-i\epsilon} 
\right\} 
\nonumber\\ &\equiv & 
\Re_0^{(2)} (t,t_0) \; e^{i\delta^0_0(t)}\, . 
\label{eq:OMNES_F2} 
\ea 
Table~\ref{table:SFF_Z} shows that with $n=2$ 
the numerical results remain indeed stable under variations of $\bar{z}$.
Moreover, as shown in Table~\ref{table:SFF}, the $O(p^2)$ approximation to
$\delta^0_0(z)$ works now much better, giving results in good agreement
with the ones obtained from the experimental phase-shifts.

Notice, that $\Omega_0^{(2)} (t,t_0)$ is not defined at 
$t_0 = 4 M_\pi^2$, because the derivative $g'(t_0)$ has a discontinuity 
at the threshold of the physical cut. 
Since  $F_S^\pi(t)$ is an analytic function in the cut s--plane, its 
Taylor expansion around the subtraction point $t_0$ ($\leq 4 M_\pi^2$) 
has a convergence 
radius $|t_0 - 4 M_\pi^2|$, which becomes zero at $t_0 = 4 M_\pi^2$. 
Thus, subtraction points close to this threshold singularity 
\cite{BCFIMS:00} should be avoided \cite{TR:00}. 
 
Since the derivative $g'(t_0)$ has been fixed at the one-loop level 
only (i.e. has been estimated at the lowest non-trivial order), 
there is a corresponding uncertainty which gets somehow 
increased by its exponentiation. This explains why the 
predicted values of $|F_S^{\pi}(M_K^2)/F_S^{\pi}(0)|$ 
in Table~\ref{table:SFF} 
are less stable for $n=2$ than for $n=1$ 
under changes of the subtraction point $t_0$. 
A twice--subtracted Omn\`es solution requires a more precise 
knowledge of the subtraction function. 
With $n=2$, the scalar form factor slightly decreases for increasing 
values of $t_0$, in the same way as $\Re_0^{(1)} (t,t_0)$, 
the tree--level once--subtracted solution, does. 
This could be easily 
improved by using the available two-loop $\chi$PT results \cite{BGT:98}. 
Nevertheless, since chiral corrections are smaller at lower 
values of $t_0$, we can safely conclude that the true value of 
$|F_S^{\pi}(M_K^2)/F_S^{\pi}(0)|$ is between 1.5 and 1.6. 
Taking the experimental phase-shift uncertainties into account, 
we finally get: 
\be\label{eq:resSFF} 
|F_S^{\pi}(M_K^2)/F_S^{\pi}(0)| \, = \, 1.55 \pm 0.10 \, . 
\ee 

The naive one-loop $\chi$PT prediction, 
$|F_S^{\pi}(M_K^2)/F_S^{\pi}(0)| = 1.61$, 
turns out to be within the $1\sigma$ range of our final 
result (\ref{eq:resSFF}). 
The advantage of using the dispersive Omn\`es resummation is that 
one can pin down the true value with an acceptable accuracy (7\%), 
in spite of having a 60\% one-loop correction.

\section{$K\to\pi\pi$ amplitudes} 
\label{K2pi}\setcounter{equation}{0} 
 
The analysis of $K\to\pi\pi$ transition amplitudes is technically 
more complicated. 
Since the electroweak scale $M_W$, where the short--distance quark 
transition takes place, is much larger than the long--distance 
hadronic scale, there are large short--distance logarithmic 
contributions which can be summed up using the 
Operator Product Expansion \cite{WI:69} 
and the renormalization group. 
One gets an effective $\Delta S=1$ Lagrangian, defined in the 
three--flavour theory \cite{GW:79,BURAS}, 
\be\label{eq:Leff} 
 {\cal L}_{\mathrm eff}^{\Delta S=1}= - \frac{G_F}{\sqrt{2}} 
 V_{ud}^{\phantom{*}}\,V^*_{us}\,  \sum_i  C_i(\nu) \; Q_i (\nu) \; ,  
 \label{eq:lag}  
\ee 
which is a sum of local four--fermion operators $Q_i$, 
constructed with the light degrees of freedom, modulated 
by Wilson coefficients $C_i(\nu)$ which are functions of the 
heavy masses $M_W$, $M_Z$, $m_t$, $m_b$ and $m_c$ 
that have been integrated out. 
The overall renormalization scale $\nu$ separates 
the short-- ($M>\nu$) and long-- ($m<\nu$) distance contributions, 
which are contained in $C_i(\nu)$ and $Q_i$, respectively. 
The physical amplitudes are of course independent of $\nu$; thus, the  
explicit scale (and scheme) dependence of the Wilson coefficients should 
cancel exactly with the corresponding dependence of the $Q_i$ 
matrix elements between on-shell states.  
We have explicitly factored out the Fermi coupling $G_F$ 
and the Cabibbo--Kobayashi--Maskawa (CKM) matrix elements 
$V_{ij}$ containing the usual Cabibbo suppression of $K$ decays. 
 
Our knowledge of $\Delta S=1$ transitions has improved qualitatively 
in recent years, thanks to the completion of the next-to-leading 
logarithmic order calculation of the Wilson coefficients 
\cite{buras1,ciuc1}. 
All gluonic corrections of $O(\alpha_s^n t^n)$ and 
$O(\alpha_s^{n+1} t^n)$ are already known, 
where $t\equiv\ln{(M_1/M_2)}$ refers to the logarithm of any ratio of 
heavy mass scales $M_1,M_2\geq\nu$.  
Moreover, the full $m_t/M_W$ dependence (at lowest 
order in $\alpha_s$) has been taken into account. 
In order to predict physical amplitudes, however, one is still 
confronted with the calculation of hadronic matrix elements of 
the four--quark operators. This is a very difficult problem, 
which so far remains unsolved. 
 
The chiral symmetry properties of the effective Lagrangian 
\eqn{eq:Leff} determine its corresponding $\chi$PT 
realization, in terms of the QCD Goldstone bosons 
\be\label{pi}  
\Phi = \left( \begin{array}{ccc}  
\sqrt{\frac{1}{2}} \pi^0 + \sqrt{\frac{1}{6}} \eta & \pi^+ & K^+ \\  
\pi^- & - \sqrt{\frac{1}{2}} \pi^0 +\sqrt{\frac{1}{6}} \eta & K^0 \\  
K^- & \bar{K}^0 & -\sqrt{\frac{2}{3}} \eta \end{array}  
\right) ,  
\ee 
parametrized through the exponential \  
$U=\exp(\sqrt{2} i  \Phi /f)$.  
At a given order in the momentum expansion, 
chiral symmetry fixes the allowed chiral operators and, therefore, 
the structure of the physical weak amplitudes. 
The only remaining problem is the calculation of the chiral couplings 
from the effective short--distance Lagrangian. 
 
The effect of strangeness--changing non-leptonic 
weak interactions with $\Delta S=1$ is incorporated \cite{CR:67} 
in the low--energy chiral theory as a perturbation to 
the strong effective Lagrangian.  
At lowest order, 
the most general effective bosonic Lagrangian, with the same 
$SU(3)_L\otimes SU(3)_R$ transformation properties as the short--distance 
Lagrangian \eqn{eq:Leff}, contains three terms: 
\ba\label{eq:lg8_g27} 
\cL_2^{\Delta S=1} &=& -{G_F \over \sqrt{2}}  V_{ud}^{\phantom{*}} V_{us}^* 
\Bigg\{ g_8 \; f^4  \;\langle\lambda L_{\mu} L^{\mu}\rangle  
 + g_{27}\; f^4 \, 
\left( L_{\mu 23} L^\mu_{11} + {2\over 3} L_{\mu 21} L^\mu_{13} 
\right)  
\Biggr. \nonumber\\ &&\qquad\qquad\quad\mbox{}   + e^2 f^6 g_{EW} \;  
\langle\lambda U^\dagger \cQ U\rangle \Bigg\} 
\, .  
\ea 
The flavour--matrix operator $L_{\mu}=-i U^\dagger D_\mu U$  
represents the octet of $V-A$ currents at lowest order in derivatives, 
$\cQ= {\rm diag}(\frac{2}{3},-\frac{1}{3},-\frac{1}{3})$ is the quark 
charge matrix, 
$\lambda\equiv (\lambda^6 - i \lambda^7)/2$ projects onto the 
$\bar s\to \bar d$ transition [$\lambda_{ij} = \delta_{i3}\delta_{j2}$] 
and $\langle A\rangle $ denotes the flavour trace of $A$. 
 
The chiral couplings $g_8$ and $g_{27}$ measure the strength of the two 
parts of the effective Lagrangian \eqn{eq:Leff} transforming as  
$(8_L,1_R)$ and $(27_L,1_R)$, respectively, under chiral rotations.   
Chiral symmetry forces the lowest--order Lagrangian to contain at least 
two derivatives (Goldstone bosons are free particles at zero momenta). 
In the presence of electroweak interactions, however, the explicit breaking 
of chiral symmetry generated by the quark charge matrix $\cQ$ induces 
the $O(p^0)$ operator $\langle\lambda U^\dagger \cQ U\rangle$ 
\cite{BW:84,GRW:86}, transforming as $(8_L,8_R)$ under the chiral group. 
In the 
usual chiral counting $e^2\sim O(p^2)$ and, therefore, the $g_{EW}$ 
term is also of order $p^2$. 
One additional term \cite{BE:85} proportional to the quark mass matrix, 
which transforms as $(8_L,1_R)$, has not been written 
since it does not contribute to the physical $K\to \pi\pi$ matrix elements 
\cite{KA91,BPP,CR:86}. 
At next--to--leading order in the chiral expansion, i.e. $O(p^4)$, a set 
of additional weak counterterms will contribute \cite{KA91,BPP,EKW:93} 
together with the strong chiral operators $L_i$ introduced in \cite{GL:85}. 
 
The Lagrangian \eqn{eq:lg8_g27} gives the lowest--order contribution 
to the $K\to 2\pi$ matrix elements. At generic values of the squared 
centre--of--mass energy $s=(p_{\pi 1}+p_{\pi 2})^2$, 
the $I=0,2$ amplitudes are given by 
\ba 
A_0(s) &=& -{G_F\over \sqrt{2}} V_{ud}V^\ast_{us}\,\sqrt{2} f\, 
\left\{ \left(g_8+{1\over 9}\, g_{27}\right) (s-M_\pi^2) 
-{2\over 3} f^2 e^2 g_{EW}\right\}  , 
\nonumber\\ 
A_2(s) &=&  -{G_F\over \sqrt{2}} V_{ud}V^\ast_{us}\, {2\over 9} f 
\, \left\{5\, g_{27}\, (s-M_\pi^2)\, - 3 f^2 e^2 g_{EW}\right\} . 
\label{TREE} 
\ea 
We have made the usual isospin decomposition: 
\ba\label{eq:AI} 
A[K^0\to \pi^+\pi^-] &\equiv& {\cal A}_0 + {1\over \sqrt{2}}\, {\cal A}_2 
\, , \nonumber\\ 
A[K^0\to \pi^0\pi^0] &\equiv&  {\cal A}_0 - \sqrt{2}\, {\cal A}_2\, , 
\\ 
A[K^+\to \pi^+\pi^0] &\equiv&  {3\over 2}\, {\cal A}_2 \, , \nonumber 
\ea 
where the amplitudes 
${\cal A}_I \equiv A_I\, \exp{\left\{i\delta_0^I\right\}}$ 
contain the strong phase-shifts, which are zero at tree level. 
 
For the discussion of the CP--conserving amplitudes we will 
neglect\footnote{ 
A general analysis of isospin breaking and electromagnetic corrections 
to $K\to 2\pi$ transitions is presented elsewhere 
\protect\cite{EIMNP:00,CDG:99,CG:00}.} 
the tiny electroweak correction proportional to $e^2 g_{EW}$. 
Taking the measured phase-shifts into account, 
eqs.~\eqn{TREE} allow us to extract the lowest--order weak 
couplings from the experimental information on $K\to 2 \pi$ decays 
\cite{PGR:86}: 
\be\label{eq:g8_g27} 
\left|g_8 + {1\over 9}\, g_{27} \right| \simeq 5.1\, , \qquad\qquad  
\vert g_{27}/ g_8 \vert \simeq 1/18 \, . 
\ee 
The huge difference between these two couplings 
shows the well--known 
enhancement of the octet $\vert\Delta I\vert = 1/2$ transitions. 
 
Let us now apply the Omn\`es procedure to the $\Delta S=1$ 
decay amplitudes. This is more subtle than for the scalar pion form 
factor, because we need to consider an off-shell kaon of 
mass squared $s=(p_{\pi 1}+p_{\pi 2})^2$, instead of a physical 
momentum transfer $s$. Since we are just studying the 
corrections induced by FSI between the two pions, the kaon can be 
formally considered as an external source, provided all 
SU(3) symmetry constraints are satisfied. 
As we saw explicitly for the scalar form factor, the FSI corrections 
that are summed up through the Omn\`es exponential 
are actually an SU(2) effect, generated by pion loops. 
Intuitively, we are just correcting a local weak $K\to\pi\pi$ 
transition with a chain of pion--loop bubbles, incorporating 
the strong $\pi\pi\to\pi\pi$ rescattering to all orders in the 
chiral expansion. 
 
In the absence of $e^2 g_{EW}$ corrections, 
the tree--level isospin amplitudes have a zero at $s=M_\pi^2$, 
because the on-shell amplitudes should vanish in the SU(3) limit 
\cite{BE:85,EIMNP:00,CA:64,GE:64}. We must take this important constraint 
into account, when making the Omn\`es summation of FSI effects, 
factorizing the zero explicitly as indicated in eq.~\eqn{eq:NSUB_ZERO}. 
This is what was done in ref.~\cite{SHORT}, using 
a once--subtracted Omn\`es factor, to evolve 
the tree-level $\chi$PT results from $s_0=M_\pi^2$ 
to the physical point $s=M_K^2$.

At higher orders in $\chi$PT there are small corrections 
proportional to $(M_K^2-M_\pi^2)$ instead of $(s-M_\pi^2)$, 
which originate in the explicit breaking of chiral symmetry 
provided by the quark mass matrix. 
According to the general one-loop analysis presented in 
appendix~\ref{ONELOOP}, those tiny effects can be neglected 
to a very good approximation. 
However, there is no need to do it. In full generality, the isospin 
amplitudes can be decomposed as 
\be\label{eq:split} 
\cA_I(s) = \tilde{a}_I(s) \left(s-M_\pi^2\right) + 
\delta \tilde{a}_I(s) \left(M_K^2-M_\pi^2\right)\, , 
\ee 
where $\delta \tilde{a}_I(s)$ is zero at lowest order\footnote{ 
To make the decomposition \protect(\ref{eq:split}) unique,
we require the function $\delta\tilde{a}_I(s)$ to depend on $s$ only 
logarithmically. }. 
Since there is a single strong phase, for a given isospin, 
the unitarity relation \eqn{eq:WASIM_EL} is also valid for the 
individual functions $\tilde{a}_I(s)$ and $\delta\tilde{a}_I(s)$. 
Therefore, the Omn\`es problem can be solved separately for the 
two pieces. 
Combining them, we can write our result for the physical 
on-shell amplitude in the simpler form: 
\ba\label{eq:OMNES_WA} 
\cA_I &\equiv &\cA_I(M_K^2) =   
\left(M_K^2-M_\pi^2\right) \; a_I(M_K^2) 
\nonumber\\ & = & 
\left(M_K^2-M_\pi^2\right) \; \Omega_I(M_K^2,s_0) \; a_I(s_0) 
\\ & = & 
\left(M_K^2-M_\pi^2\right) \; \Re_I(M_K^2,s_0) \; a_I(s_0) 
\; e^{i\delta^I_0(M_K^2)}\, , 
\nonumber\ea 
where $a_I(s)\equiv \tilde{a}_I(s) + \delta\tilde{a}_I(s)$. 
 
The once--subtracted Omn\`es factor $\Omega_I^{(1)}(M_K^2,s_0)$ 
is universal, 
i.e. the same for $f(s) = F_S^\pi(s)$, $\tilde{a}_I(s)$ or 
$\delta\tilde{a}_I(s)$, 
because it only depends on the strong phase-shift 
$\delta^I_0(s)$ [see eq.~\eqn{eq:OMNES_F}]. 
This is no longer true with two subtractions [eq.~\eqn{eq:OMNES_F2}], 
because $\Omega_I^{(2)}(s,s_0)$ contains an explicit dependence on 
$f'(s_0)/f(s_0)$. 
Nevertheless, given the smallness of the non-leading 
$\delta\tilde{a}_I(s)$ contribution, 
it is a very good numerical approximation to take also 
a global Omn\`es exponential for $a_I(s)$ in the 
twice-subtracted case. 
 
Let us define $a_0(s)\equiv a_0^{(8)}(s) + a_0^{(27)}(s)$, 
thus separating the $(8_L,1_R)$ and $(27_L,1_R)$ contributions 
to the isoscalar amplitude. The complete one-loop $\chi$PT results 
for the different decay amplitudes are given in 
appendix~\ref{ONELOOP}. Their $s$ dependences 
can be written in a rather transparent way: 
\ba\label{eq:s_dep} 
a_0^{(8)}(s) &=& a_0^{(8)}(0)\;\left\{ 1 + g_0^{(8)}(s) 
  + O(p^4)\right\}\, , 
\nonumber\\ 
a_0^{(27)}(s) &=& a_0^{(27)}(0)\;\left\{ 1 + g_0^{(27)}(s) 
  + O(p^4)\right\}\, , 
\\ 
a_2(s) &=& a_2(0)\;\left\{ 1 + g_2(s) 
  + O(p^4)\right\}\, , 
\nonumber\ea 
where 
\ba\label{eq:g_0_2} 
g_0^{(8)}(s) &=& {s\over f^2}\left\{ \left (1-{M_\pi^2\over 2s}\right ) 
\bar{J}_{\pi\pi}(s) - {1\over 4}\,\left(1-{M_K^2\over s}\right) 
\bar{J}_{KK}(s) + {M_\pi^2\over 18 s}\,\bar{J}_{\eta\eta}(s)  
\right .\nonumber\\ 
&&\left.\quad\mbox{} +C^8_5(\mu ) + {3\over 4(4\pi )^2} 
\left (\ln{\mu^2\over M_\pi^2} -1\right ) +  {1\over 4(4\pi )^2} 
\ln{M_K^2\over M_\pi^2} \right\}\, , 
\\ 
g_0^{(27)}(s) &=& {s\over f^2}\left\{ \left (1-{M_\pi^2\over 2s}\right ) 
\bar{J}_{\pi\pi}(s) - {3\over 2}\,\left(1-{M_K^2\over s}\right) 
\bar{J}_{KK}(s) - {M_\pi^2\over 2 s}\,\bar{J}_{\eta\eta}(s)  
\right .\nonumber\\ 
&&\left.\quad\mbox{} + C^{27}_5(\mu) - {1\over 2(4\pi )^2} 
\left (\ln{\mu^2\over M_\pi^2} -1\right ) +  {3\over 2(4\pi )^2} 
\ln{M_K^2\over M_\pi^2} \right\}\, , 
\\ 
g_2(s) &=& {s\over f^2}\left\{ -\frac{1}{2} 
\left (1-{2 M_\pi^2\over s}\right ) \bar{J}_{\pi\pi}(s) 
\right .\nonumber\\ &&\left.\quad\mbox{} 
+ \bar{C}^{27}_5(\mu) - {1\over 2(4\pi )^2} 
\left (\ln{\mu^2\over M_\pi^2} -1\right )  \right\}\, . 
\ea 

The corrections coming from the $\delta\tilde{a}_I$ terms, 
included in these results, are very small. 
Denoting by $\tilde{g}_I(s)$ the corresponding functions 
for the uncorrected $\tilde{a}_I$ amplitudes, the 
differences $\Delta g_I(s)\equiv {g}_I(s) - \tilde{g}_I(s)$ 
only get contributions 
from the $K\bar K$ and $\eta\eta$ loop functions which, 
as shown in section~\ref{Scalar}, are numerically suppressed at low values  
of s. Moreover, they get an additional suppression factor $M_\pi^2$ 
(see appendix B). 
Since the $K\bar K$ and $\eta\eta$ intermediate states cannot give 
rise to $I=2$, one gets $\Delta g_2(s) = 0$ 
($\delta\tilde{a}_2$ does not have any $s$ dependence at this order), 
while the isoscalar differences are given by 
$\Delta g^{(8)}_0(s)= 
M_\pi^2\, [9 \bar{J}_{KK}(s) +8 \bar{J}_{\eta\eta}(s)]/(36 f^2)$ 
and 
$\Delta g^{(27)}_0(s) = 
M_\pi^2\, [3 \bar{J}_{KK}(s) - 4 \bar{J}_{\eta\eta}(s)]/(2 f^2)$. 
Even at $s=M_K^2$, these differences are completely negligible: 
$\Delta g^{(8)}_0(M_K^2) \sim 1 \times 10^{-3}$ and 
$\Delta g^{(27)}_0(M_K^2) \sim - 2\times 10^{-4}$.

Notice the strong similarity with the scalar form factor result 
in eq.~(\ref{eq:g_scalar}). The isoscalar functions 
$g^{(8)}_0(s)$ and $g^{(27)}_0(s)$ get exactly the same 
$\bar{J}_{\pi\pi}(s)$ contribution than the scalar form 
factor function $g(s)$, while the corresponding contribution to 
$g_2(s)$ has opposite sign. The polynomial factors 
in front of the $\pi\pi$ loop function, $(s - M_\pi^2/2)$ 
and $(M_\pi^2 - s/2)$ for $I=0$ and 2 respectively, clearly 
identify the corresponding lowest--order $\chi$PT phase-shifts, 
given in eq.~(\ref{eq:CHPT_delta}) 
[Im$\,\bar{J}_{\pi\pi}(s) = \theta(s-4M_\pi^2)\,\sigma_\pi(s)/(16\pi)$]. 
 
In addition, at $\mu^2=M_K^2$, one recognizes the same local 
infrared logarithmic enhancement of all isoscalar $g$ functions, 
$\ln{(M_K^2/M_\pi^2)}/(4\pi)^2$; 
the corresponding factor in $g_2$ decreases the $I=2$ amplitude. 
For arbitrary values of the chiral scale $\mu$, 
this logarithmic correction is split in 
$\ln{(\mu^2/M_\pi^2)}$ and $\ln{(M_K^2/M_\pi^2)}$ terms, 
which are slightly different for the different $g$ functions. 
However, all isoscalar $g$ functions contain exactly the 
same infrared $\ln{M_\pi^2}$ contribution.

 
Thus, the $s$ dependence of the weak decay amplitudes is indeed 
dominated by infrared effects related to the FSI of the two 
final pions. Moreover, in the isoscalar case, by comparison with the  
scalar form factor, we see explicitly 
that this is a universal effect related to the quantum numbers 
of the $\pi\pi$ state. 
 
The particular dynamics leading to this final state gives rise also 
to local contributions, which are different in 
each case. We saw in section~\ref{Scalar} that at the usual 
reference scale $\mu=M_\rho$ the contribution from the local 
term $[L_5^r + 2 L_4^r](M_\rho)$ is small. For the weak 
amplitudes this needs to be further investigated; the usual 
factorization models \cite{EKW:93} amount to 
$C^8_5(M_\rho) = C^{27}_5(M_\rho) = \bar{C}^{27}_5(M_\rho) = 0$. 
 
The main effects of the short--distance dynamics, 
not related to FSI, are 
contained in the particular values of the different amplitudes $a_I(s)$ at 
$s=0$. This physics needs to be analyzed independently, because 
it cancels out from the Omn\`es relation. 
The Omn\`es factor only allows us to relate the amplitudes at 
two different values of $s$, but does not give any information 
on their global normalization.

Taking a low subtraction point where higher--order 
corrections are expected to be small, 
we can just multiply the tree--level 
formulae \eqn{TREE} with the experimentally determined Omn\`es 
exponentials, as done in ref.~\cite{SHORT}. 
For $I=0$ we already have the result obtained in the previous 
section, 
\be\label{eq:R0} 
\Re_0(M_K^2,0) = 1.55 \pm 0.10 \, , 
\ee 
which improves the lowest-order estimates made 
in refs.~\cite{SHORT,TR88} at $s_0 = M_\pi^2$. 
 
In the $I=2$ channel the inelasticity effect is absent at least up to 
1.6 GeV. Evaluating the once-subtracted dispersive integral over 
the measured phase-shifts up to $\bar z =(1.6$ GeV)${}^2$, we get 
\be\label{eq:R2} 
\Re_2^{(1)}(M_K^2,0) = 0.92 \pm 0.03 \, ,  
\ee 
to be compared with the earlier estimates 
$\Re_2^{(1)}(M_K^2,M_\pi^2) = 0.96$ ($\bar z =1$~GeV${}^2$) \cite{ME:91} and 
$\Re_2^{(1)}(M_K^2,M_\pi^2) = 0.89 \pm 0.03$ \cite{NS:91}. 
The error bar in (\ref{eq:R2}) takes into account uncertainties in the  
fits to the phase--shift data and higher energy contributions. 
 
The corrections induced by FSI in the moduli of the decay amplitudes 
${\cal A}_I$ generate an additional enhancement of the 
$\Delta I=1/2$ to $\Delta I=3/2$ ratio, 
\be\Re_0(M_K^2,0)/\Re_2(M_K^2,0) = 1.68\pm 0.12 \, .\label{eq:ratio}\ee 
This factor multiplies the enhancement already found at short distances. 
This is a quite large correction, 
which improves previous calculations of $A_I(M_K^2)$. 
Taking the $\Re_I$ correction into account, the experimental 
$A_I$ amplitudes imply the following corrected values for the 
lowest--order $\Delta S=1$ chiral couplings: 
\be 
\left|g_8 + {1\over 9}\, g_{27} \right| \approx 3.3 \qquad , \qquad 
|g_{27}| \approx 0.31 \, . 
\ee 
These ``experimental'' numbers are not very far from the 
short--distance estimates obtained in the first of refs.~\cite{PR91}.

\section{Standard Model prediction of $\varepsilon'/\varepsilon$} 
\label{eps}\setcounter{equation}{0} 
 
One further subtlety has to be taken into account in the discussion 
of CP--violating isospin amplitudes. 
The derivation of eq.~(\ref{eq:ASIM}) for the absorptive 
parts makes use of Time--Reversal invariance, so that 
the procedure can be strictly applied only to CP--conserving amplitudes. 
This is not a problem, however, because we are working to first 
order in the weak Fermi coupling. 
 
The CP--odd phase is hidden in the Wilson coefficients of 
the short--distance $\Delta S=1$ Lagrangian \eqn{eq:Leff}, 
which can be decomposed as 
\be 
C_i(\nu ) = z_i(\nu ) + \tau\, y_i(\nu ) \qquad ; \qquad 
\tau = -{ V_{td}^{\phantom{*}}\,V^*_{ts} \over 
        V_{ud}^{\phantom{*}}\,V^*_{us}} \, . 
\ee 
Since CP violation is only originated by the short--distance 
ratio of CKM matrix elements $\tau$, we can always write 
\be 
\cA_I = \cA_I^{CP} + \tau \,\cA_I^{\slashchar{CP}} 
\ee 
and apply the Omn\`es procedure to the 
amplitudes $\cA_I^{CP}$ and $\cA_I^{\slashchar{CP}}$, 
which respect Time--Reversal invariance. 
 
The CP-conserving piece of $\tau \,\cA_I^{\slashchar{CP}}$ is 
negligible in comparison with $\cA_I^{CP}$. Therefore, in a 
more standard notation, 
$\mbox{Re}A_I \approx A_I^{CP}$ and 
$\mbox{Im}A_I = \mbox{Im}(\tau )\, A_I^{\slashchar{CP}}$, 
where ``real'' and ``imaginary'' refer to CP--even and CP--odd 
since the absorptive phases have been already factored out 
through $\cA_I = A_I\, e^{i\delta_0^I}$.

The most striking consequence of the correction factors $\Re_{0,2}$ 
is a sizeable modification of the numerical short--distance 
estimates\footnote{ 
The correction factors $\Re_I^{(1)}(M_K^2,M_\pi^2)$ 
were already considered in ref.~\protect\cite{FGYOPR:92} 
to estimate $\epsilon'/\epsilon$ within the 
$SU(2)_L\otimes SU(2)_R\otimes U(1)$ model of CP violation.} 
for the direct CP--violation 
parameter $\varepsilon^\prime/\varepsilon$. 
A handy way of writing this quantity, 
used in all theoretical short--distance calculations up to date, can be as 
follows \cite{Buras} 
\begin{equation} 
{\varepsilon^\prime\over\varepsilon} = 
\mbox{Im}\left( V_{ts}^* V_{td}^{\phantom{*}}\right) 
\; e^{i\Phi}\; 
\left [P^{(1/2)} - P^{(3/2)}\right ]\, , 
\label{EPS} 
\end{equation} 
where 
the phase $\Phi = \Phi_{\varepsilon^\prime} -  \Phi_\varepsilon \simeq 0$ 
and the quantities $P^{(1/2)}$ and $ P^{(3/2)}$ contain the contributions 
from the hadronic matrix elements of four--quark operators 
with $\Delta I =1/2$ and $3/2$ respectively: 
\begin{eqnarray} 
P^{(1/2)}&=& r \,\sum_i y_i(\nu)\, \langle Q_i(\nu)\rangle_0 \, 
(1-\Omega_{IB}) \, , 
\nonumber\\ 
 P^{(3/2)}&=& {r\over \omega} \,\sum_i y_i(\nu)\, \langle Q_i(\nu)\rangle_2 
\, . \label{P_I} 
\ea 
Here, 
$\langle Q_i\rangle_I\equiv \langle (\pi\pi )_I\vert Q_i\vert K\rangle$, 
$r$ and $\omega$ are given by 
\be 
r= {G_F\over 2 \vert\varepsilon\vert}\, {\omega\over \mbox{Re}A_0}\, 
\qquad , \qquad 
\omega = {\mbox{Re}A_2\over \mbox{Re}A_0}\, , 
\label{eq:r_omega} 
\ee 
and the parameter 
\be 
\Omega_{IB} = {1\over \omega} 
{(\mbox{Im}A_2)_{IB} 
\over \mbox{Im}A_0} 
\label{eq:isospin} 
\ee 
parametrizes isospin breaking corrections. 
 
A detailed analysis of $\epsilon'/\epsilon$, within the Standard 
Model, will be given in ref.~\cite{PPS:00}. Here we just want to 
illustrate the important role of FSI and how their proper inclusion 
modifies the $\epsilon'/\epsilon$ prediction in a very important way.

Since the hadronic matrix elements are quite uncertain theoretically, 
the CP--conserving amplitudes $\mbox{Re}A_I$, and thus 
the factors $r$ and $\omega$, are set to their experimentally 
determined values; this automatically includes the FSI effect. 
All the rest in the numerator is {\em theoretically} predicted via 
short--distance calculations, because the leading contributions come 
\cite{Trieste} 
from the operators $Q_6$ and $Q_8$ whose matrix elements cannot 
be directly measured from $K\to 2\pi$ decay rates. 
 
As a consequence, since the relevant matrix elements 
$\langle Q_{6,8}\rangle_I$ 
are usually taken from large--$N_C$ estimates \cite{Buras} or 
lattice calculations \cite{Lattice}, 
which do not include FSI corrections, 
this procedure produces a mismatch with the FSI 
included phenomenologically in the values of $r$ and $\omega$. 
This can be easily corrected, introducing in the numerator 
the dispersion factors $\Re_I$ that we have estimated. 
This implies \cite{SHORT}
a large enhancement of the predicted value of 
$\varepsilon'/\varepsilon$ by roughly a factor of 2. 
 
To a very good approximation, the Standard Model prediction for 
$\varepsilon'/\varepsilon$ can be written, up to global factors, 
as \cite{Buras} 
\be 
{\varepsilon'\over\varepsilon} \sim 
\left [ B_6^{(1/2)}(1-\Omega_{IB}) - 0.4 \, B_8^{(3/2)} 
 \right ]\, , 
\label{EPSNUM} 
\ee 
where $B_6^{(1/2)}$ and $B_8^{(3/2)}$ parametrize the matrix elements 
of the QCD penguin operator $Q_6$ and the electroweak penguin operator $Q_8$, 
respectively, in units of their vacuum insertion approximation values. 
These parameters are usually taken to be  
(from $1/N_C$ considerations \cite{Buras} and  
Lattice calculations \cite{Lattice}) 
$B_6^{(1/2)}= 1.0\pm 0.3$ and $B_8^{(3/2)} = 0.8\pm 0.2$, while 
the isospin--breaking  factor is set to 
$\Omega_{IB}\approx 0.25$ \cite{Omega} 
with large uncertainties \cite{GV:99,MW:00}. With those inputs, there is 
a rather large numerical cancellation between the two terms in 
eq.~\eqn{EPSNUM}, which results in a 
predicted central value \cite{Buras,Lattice} 
$\varepsilon'/\varepsilon \approx 7.0\times 10^{-4}$. 
 
Since those estimates do not include FSI effects, their values should 
be multiplied by the appropriate factors $\Re_I$. Notice, that the 
Omn\`es procedure can be also applied to the individual matrix elements 
$\langle Q_i\rangle_I$. 
In order to avoid 
any possible double counting, we will take as the starting point of 
our analysis the large--$N_C$ estimate for the relevant matrix elements 
\cite{BURAS}: 
\be\label{eq:B_6_8_Nc} 
\left. B_6^{(1/2)}\right|_{N_C\to\infty} \, =  \, 1 
\; ; \qquad\quad 
\left. B_8^{(3/2)}\right|_{N_C\to\infty}\, \approx \, 1.0 \; . 
\ee 
FSI only appear at next-to-leading order in the $1/N_C$ expansion and, 
therefore, correct the leading values \eqn{eq:B_6_8_Nc}. 

The corrected $\varepsilon'/\varepsilon$ prediction can be 
easily obtained, taking into account the following points: 
 
\begin{enumerate} 
 
\item 
The penguin operator $Q_6$ transforms as $(8_L,1_R)$ under chiral 
transformations. At lowest order in the chiral expansion, it corresponds 
to the first operator in eq.~\eqn{eq:lg8_g27} (from 
eq.~\eqn{eq:B_6_8_Nc} one actually gets the $Q_6$ contribution to the 
chiral coupling $g_8$, in the large--$N_C$ limit). 
The FSI corrections induced by pion chiral loops modify $B_6^{(1/2)}$
as follows
$$ 
B_6^{(1/2)} \, = \, \left. B_6^{(1/2)}\right|_{N_C\to\infty} \; 
\times\;\Re_0(M_K^2,0) \, = \, 1.55\; . 
$$ 
 
\item 
The electroweak penguin operator $Q_8$ corresponds to the chiral 
operator proportional to $g_{EW}$ in eq.~\eqn{eq:lg8_g27}. 
As shown in \eqn{TREE}, it contributes to the two isospin amplitudes, 
although we only need here the $I=2$ piece $\langle Q_8\rangle_2$. 
The $K\to 2\pi$ matrix element is not proportional to 
$M_K^2-M_\pi^2$ because the needed $SU(3)$ breaking is provided 
by the quark charge matrix (the chiral operator is identically 
zero for $\cQ=I$). The presence or not of this factor does not change 
the Omn\`es summation (the corresponding zero just factors out whenever 
is present). One gets then 
$$ 
B_8^{(3/2)} \, =\, \left. B_8^{(3/2)}\right|_{N_C\to\infty} \; 
\times\;\Re_2(M_K^2,0) 
 \, = \, 0.92\; . 
$$ 
 
\item
The isospin--breaking correction coming from $\pi^0$--$\eta$ mixing has been 
recently calculated at $O(p^4)$ in the chiral expansion, with the result 
$\Omega_{IB}= 0.16\pm 0.03$ \cite{EMNP:00}. 
This value is smaller than the previous lowest--order estimate 
$\Omega_{IB}\approx 0.25$  \cite{Omega}. 
The term $B_6^{(1/2)}\Omega_{IB}$ in eq.~(\ref{EPSNUM}) 
should be multiplied by  $\Re_2$ and not by $\Re_0$, 
because it corresponds to two final pions with $I=2$. Thus, 
$$ 
B_6^{(1/2)}\;\Omega_{IB} \, =\,  \left. B_6^{(1/2)}\right|_{N_C\to\infty}
\;\Omega_{IB}
\times\;\Re_2(M_K^2,0) \, = \, 0.15\; . 
$$ 

\end{enumerate} 
 
The large FSI correction to the $I=0$ amplitude gets reinforced 
by the mild suppression of the $I=2$ contributions. The net effect 
is a large enhancement of eq.~\eqn{EPSNUM}, by a factor 2.4, 
pushing the predicted central value from $7.0\times 10^{-4}$ 
\cite{Buras,Lattice} to 
\be\label{eq:simpvalue} 
\varepsilon'/\varepsilon = 17\times 10^{-4}\, , 
\ee 
which compares well with the present experimental world average 
\cite{Ceccuchi} in eq.~\eqn{eq:exp}. 

A more careful analysis, taking into account all hadronic and 
quark--mixing inputs \cite{PPS:00} gives the result quoted in 
eq.~\eqn{eq:epsth} for the Standard Model prediction of 
$\varepsilon'/\varepsilon$.

\section{Discussion} 
\label{CONC}\setcounter{equation}{0} 
 
Many attempts have been made to compute the isospin amplitudes $A_I$ 
from first principles \cite{BBG_1,Buras,Lattice,Trieste,Dortmund,PGR:86,
PR91,JP94,BBG87,KPR98,BP95,NA:00,BBLM:99,HYCH:99}. 
Although those calculations have provided encouraging results, we 
are still far from getting accurate predictions. Nevertheless, a 
qualitative understanding of the $K\to\pi\pi$ transition amplitudes 
is now emerging. 

The strong rescattering of the two final pions generates important 
corrections to the kaon decay amplitudes, 
enhancing the $I=0$ piece by about 50\% and originating a mild suppression 
of the $I=2$ one.
FSI alone cannot explain the measured ratio of $\Delta I=1/2$ to 
$\Delta I=3/2$ transition amplitudes, but they 
constitute a very important ingredient which reinforces the enhancement 
already found at short distances. 
Combined with the $1/N_C$ and $\chi$PT expansions,  
the calculation of the 
Omn\`es factors $\Omega_{0,2}(M_K^2,0)$ allows for a reliable estimate 
of $\varepsilon'/\varepsilon$ \cite{SHORT,PPS:00}. 
 
The lowest--order approximation in the $1/N_C$ expansion does not 
provide a good starting point to analyze the CP--conserving $K\to 2\pi$ 
amplitudes, because the anomalous dimensions of the most important 
operators $Q_i$ are zero at this order \cite{PR91}. 
Thus, at lowest order in $1/N_C$ 
one misses the dominant physics leading to the well--known 
short--distance enhancement. That makes difficult to perform precise 
predictions for the $K\to 2\pi$ decay rates. 
 
The situation is different for the CP--violating amplitudes, which 
are completely dominated by $Q_6$ and $Q_8$. These are precisely the 
only operators which have a non-zero anomalous dimension at leading 
order in the $1/N_C$ expansion. The large--$N_C$ approximation 
works rather well for those operators \cite{PR91,JP94} 
and their matrix elements can be 
safely estimated within a 30\% accuracy, once the large infrared 
logarithms related to FSI are properly taken into account. 
 
In the large--$N_C$ limit the four--quark operators factorize into 
currents which have a known chiral realization at very low energies. 
The factorization of the operators $Q_6$ and $Q_8$ leads to 
scalar (pseudo-scalar) currents which are not directly measurable; 
their matrix
elements are determined with $\chi$PT techniques at leading 
(next-to-leading for $Q_8$) 
non-trivial order in the momentum expansion. 
This fixes the $Q_6$ and $Q_8$ contribution to the $\Delta S=1$ 
$\chi$PT couplings in the large--$N_C$ limit \cite{SHORT,PPS:00}. 
A reliable determination of the corresponding 
$K\to 2\pi$ transition amplitudes can then be performed 
at low $s$ values where chiral loop corrections 
are smaller. Once this is accomplished, the Omn\`es dispersive factors 
allow us to evolve this result to the physical $s=M_K^2$ point, 
resumming the large chiral corrections associated with FSI. 
 
The usual vacuum insertion estimate of $\langle Q_6\rangle_0$, 
adopted in some {\it Standard Model} calculations of 
$\varepsilon'/\varepsilon$ \cite{Buras,Lattice}, 
corresponds to the lowest non-trivial order in both the 
$1/N_C$ and $\chi$PT expansions. 
This naive estimate misses the large enhancement generated by 
one-loop $\chi$PT corrections 
\cite{KA91,BPP,JHEP}, which originates mainly in the strong 
rescattering of the two final pions with $I=J=0$. 
The FSI correction destroys the accidental numerical 
cancellation between the $Q_6$ and $Q_8$ contributions in 
eq.~(\ref{EPSNUM}), producing a large increase in the 
resulting prediction of $\varepsilon'/\varepsilon$.
The size of the FSI effect can be already determined with
the one-loop $\chi$PT calculation. 
The Omn\`es resummation is only needed to perform a 
reliable estimate of higher--order corrections and pin down 
their associated uncertainties.
 
More work is still needed in order to get a precise quantitative 
description of kaon decays. In the meanwhile, our analysis 
demonstrates that it is at least possible to pin down the value of 
$\varepsilon'/\varepsilon$ with an accuracy of about 30\%. 
Within the present uncertainties, the resulting 
Standard Model theoretical 
prediction is in good agreement with the measured experimental 
value, without any need to invocate a new physics source of 
CP violation. 
 
\section*{Acknowledgments}
 
We are grateful to I.~Scimemi for his many comments on this work. 
We would also like to acknowledge discussions with A.J.~Buras, 
G.~Colangelo, J.~Gasser, V.~Gim\'enez, M.~Golterman,
G.~Isidori, M.~Jamin, A.~Manohar, G.~Martinelli,  
H.~Neufeld, J.A.~Oller, E.~Oset, J.~Prades, J.~Portol\'es, E. de Rafael 
and L.~Silvestrini. This work has been supported in part by 
the European Union TMR Network EURODAPHNE (Contract No. 
ERBFMX-CT98-0169), and by DGESIC (Spain) under grant No. PB97-1261. 
EP is supported by the Ministerio de Educaci\'on  y Cultura (Spain).

\appendix 
\renewcommand{\theequation}{\Alph{section}.\arabic{equation}} 
\setcounter{section}{0}\setcounter{equation}{0}

\section{Experimental $\pi$--$\pi$ phase shifts} 
\label{App:phaseshift}\setcounter{equation}{0} 
 
For the experimental phase shifts of $\pi$--$\pi$ scattering with $J=0$ and 
total isospin $I=0$ or 2 we used a simple parametrization by A. Schenk 
\cite{SCHENK} that works in the elastic region. A more involved 
coupled--channel 
analysis is needed above the first inelastic threshold \cite{MANY}. 
For generic $I$ and $J$, the parametrization provided in \cite{SCHENK} 
is given by: 
\ba 
\label{PHASE} 
\tan\delta_J^I(s) &=& \sqrt{ {s-4M_\pi^2\over s}} \; 
\left ({s-4M_\pi^2\over 4M_\pi^2}\right )^J  \; 
\left ({4M_\pi^2-s_J^I\over s-s_J^I}\right ) 
\nonumber\\ 
&\times & \left\{  a_J^I + \tilde{b}_J^I 
\left ({s-4M_\pi^2\over 4M_\pi^2}\right ) 
+c_J^I\left ({s-4M_\pi^2\over 4M_\pi^2}\right )^2 \right \} 
\, . 
\ea 
The threshold expansion of the scattering amplitude is reproduced by setting 
\be 
\tilde{b}_J^I = b_J^I -a_J^I{4M_\pi^2\over s_J^I-4M_\pi^2} + 
(a_J^I)^3\delta_{J0}\, . 
\ee 
For given $I,J$ there are four parameters: $a_J^I,\, b_J^I,\, c_J^I$ and 
$s_J^I$. The numerical values of $a_J^I,\, b_J^I$ (which parametrize the 
threshold behaviour of the scattering amplitude) have been determined by 
means of $\chi$PT, while the remaining two parameters have been extracted 
from the experimental data.  
The details of the analysis can be found in the original work \cite{SCHENK}. 
Here we only compile the numerical values of the parameters. 
For isospin $I=0$ and $I=2$, in S--wave, the values of the threshold 
parameters are: 
\be\label{I0} 
a_0^0 = 0.20 \quad ; \quad b_0^0 = 0.24  \quad ; \quad 
a_0^2 = -0.042 \quad ; \quad  b_0^2 = -0.075  \, . 
\ee 
For the other parameters $ c_J^I$ and $s_J^I$ we have taken the range 
of values 
determined in \cite{SCHENK} in order to take into account the experimental 
uncertainties. 
For $I=0$, we used $c_0^0 = 0.008,\, 0.0,\, -0.015$ and 
$\sqrt{s_0^0} =840,\, 865,\, 890$ MeV.  
For $I=2$, we used $c_0^2 = 0$ and 
${s_0^2} = -920^2,\, -685^2,\, -555^2$ MeV${}^2$. 
The three values for each single parameter correspond to the three
 solid curves 
shown in Figure~\ref{fig:PHASESHIFT} for each isospin. 
The central line corresponds to the best fit in \cite{SCHENK}, while the 
other two extremes enclose the region covered by the experimental data 
considered in \cite{SCHENK}. 
 
\begin{figure}[htb] 
\centerline{\mbox{\epsfxsize=10cm\epsffile{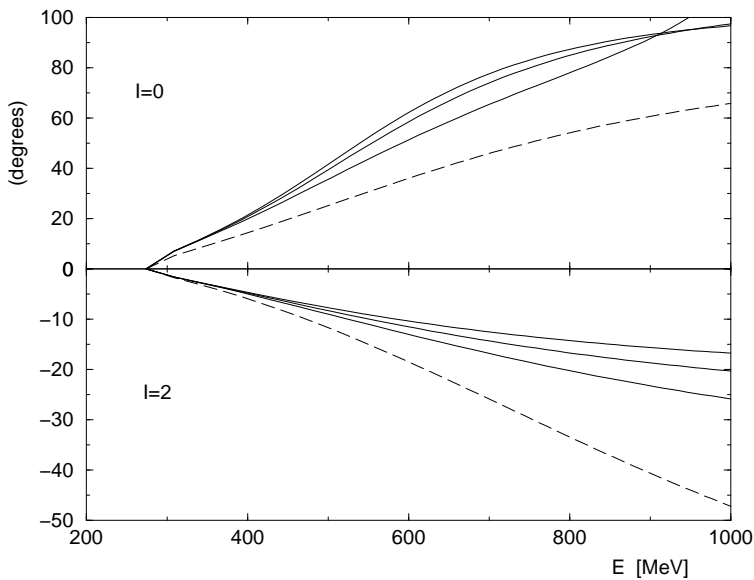}}} 
\caption{Phase shifts $\delta_0^{0;2}(s)$ used in our numerical analyses. 
The solid lines enclose the range covered by the experimental data, while 
the dashed lines show the unitarized lowest--order $\chi$PT prediction.} 
\label{fig:PHASESHIFT} 
\end{figure} 

The lowest--order $\chi$PT prediction of the 
phase shifts is also shown with a dashed line in Figure~\ref{fig:PHASESHIFT}. 
This corresponds to the expression 
\be\label{eq:CHPT_delta} 
\tan\delta_0^{0;2}(s) = {1\over 32\pi f^2} \; \sigma_\pi (s)\;\left ( 
  2s-M_\pi^2\, ;\,  2M_\pi^2-s\right )\,;~~~~~ 
\sigma_\pi (s)\equiv \sqrt{1-{4M_\pi^2\over s}}, 
\ee 
which is a unitarization of the usual $\chi$PT prediction, valid at 
low values of $s$ where $\tan\delta\sim \delta$. 
The lowest--order $\chi$PT prediction 
fails already at relatively low energies $\sim 500$ MeV, 
specially for $I=0$. 
In the $I=0$ case it underestimates 
the experimental phase shift, while in the $I=2$ 
case it gives a too large, in absolute value, phase shift.

\section{$K\to\pi\pi$ matrix elements at one loop} 
\setcounter{equation}{0} 
\label{ONELOOP}

The one-loop contribution to the physical CP--conserving 
$K\to\pi\pi$ isospin amplitudes has been computed in Refs. \cite{BPP,KA91}. 
In this appendix the same amplitudes are calculated at a generic value 
of the squared invariant mass $s = (p_{\pi 1}+p_{\pi 2})^2$. 
The complete next--to--leading correction is of order $p^4$ in the chiral 
expansion and includes one-loop contributions generated by the 
lowest--order $p^2$ Lagrangian (\ref{eq:lg8_g27}) 
and tree--level contributions 
coming from order $p^4$ counterterms \cite{KA91,BPP,EKW:93}. 
 
In the analysis of the CP--conserving amplitudes we have neglected the 
tiny electroweak corrections which are proportional to $e^2 g_{EW}$ 
at leading order in the chiral expansion. 
We then decompose the isospin amplitudes as follows 
\be\label{eq:split_app} 
\cA_I(s) = \tilde{a}_I(s) \left(s-M_\pi^2\right) + 
\delta\tilde{a}_I(s) \left(M_K^2-M_\pi^2\right)\, , 
\ee 
where $\delta\tilde{a}_I(s)$ is zero at lowest order. 
In addition, we define 
$\tilde{a}_0\equiv \tilde{a}_0^{(8)}+ \tilde{a}_0^{(27)}$ 
and $\delta\tilde{a}_0\equiv \delta\tilde{a}_0^{(8)}+ 
\delta\tilde{a}_0^{(27)}$, thus explicitly 
separating the octet and 27--plet contributions to the $I=0$ amplitude. 
 
At $O(p^4)$ the octet $I=0$ function $\tilde{a}_0^{(8)}$ takes the form 
 
\ba 
\label{ONELOOP_8} 
&&\hspace{-0.8cm} \tilde{a}_0^{(8)}= 
-{G_F\over \sqrt{2}} V^{\phantom{\ast}}_{ud}V^\ast_{us} 
 \sqrt{2} f_\pi \, g_8 \;\Biggl\{ 1 
- {1\over 2} \left( 2 s - M_\pi^2 \right) B(M_\pi^2,M_\pi^2,s) 
\Biggr. \nonumber\\ 
&&\hspace{-0.8cm} \mbox{} 
+ {1\over 4} \left( s - M_K^2 + M_\pi^2\right) B(M_K^2,M_K^2,s) 
+{1\over 6}  M_\pi^2\, B(M_\eta^2,M_\eta^2,s) 
\nonumber\\ 
&&\hspace{-0.8cm} \mbox{} 
+  {(M_K^2-M_\pi^2)\over  12 M_\pi^2} \left[ 
  ( 3 M_K^2 - 4 M_\pi^2 )\, B(M_K^2,M_\pi^2,M_\pi^2 ) 
  + M_K^2\, B(M_K^2,M_\eta^2,M_\pi^2 ) \right] 
\nonumber\\ 
&&\hspace{-0.8cm} \mbox{} 
+ {67\over 12}\,\mu_\pi + {1\over 6}\,\mu_K -{7\over 36}\,\mu_\eta 
+{(M_K^2-M_\pi^2)\over 2 M_\pi^2}  \left (\mu_\pi-\mu_\eta\right ) 
-{M_\pi^2\over 2 M_K^2}\left (\mu_\pi + {\mu_\eta\over 9} \right ) 
\nonumber\\ 
&&\hspace{-0.8cm} \mbox{} + 
 {1\over f^2} \biggl[ 
-16 (2M_K^2+M_\pi^2)\, L_4^r(\mu)  -4 (M_K^2+3M_\pi^2)\, L_5^r(\mu) 
+ M_K^2\, C^8_1(\mu) 
\biggr. \nonumber\\ && \Biggl. \biggl.\mbox{} 
+ M_\pi^2\, C^8_4(\mu) 
- {M_\pi^4\over M_K^2}\, C^8_6(\mu) + s\, C^8_5(\mu) \biggr] 
 \Biggr\}\, , 
\ea 
 
\noindent while $ \delta\tilde{a}_0^{(8)}$ is given by 
 
\ba 
\label{ONELOOP_8_D} 
&&\hspace{-0.8cm} \delta\tilde{a}_0^{(8)}= 
 -{G_F\over \sqrt{2}} V^{\phantom{\ast}}_{ud}V^\ast_{us} 
 \sqrt{2} f_\pi \, g_8 \; \Biggl\{ 
 - M_\pi^2\left[{1\over 4}B(M_K^2,M_K^2,s) + {2\over 9} 
B(M_\eta^2,M_\eta^2,s) \right] 
 \Biggr.   \nonumber\\ 
&&\hspace{-0.8cm} \mbox{} 
-{1\over 12} (5 M_K^2+ 4 M_\pi^2)\, B(M_K^2,M_\pi^2,M_\pi^2) 
+{1\over 12} M_K^2\,  B(M_K^2,M_\eta^2,M_\pi^2) 
 \nonumber\\ 
&&\hspace{-0.8cm} \mbox{} 
+ {1\over f^2}\left[  M_\pi^2\, C^8_2(\mu) 
  + M_K^2\, C^8_3(\mu)  +{M_\pi^4\over M_K^2} C^8_6(\mu) \right] 
    -{5\over 6}\,\mu_\pi-{\mu_\eta\over 18} +{4\over 3}\,\mu_K 
\nonumber\\ 
&& \hspace{-0.8cm} \mbox{} 
\Biggl. -{M_\pi^2\over M_K^2 -M_\pi^2} 
\left ( {\mu_\pi\over 2} +{17\over 18}\mu_\eta -2\mu_K 
 \right) 
     -{M_\pi^4\over 2 M_K^2(M_K^2 -M_\pi^2)}\left( \mu_\pi 
+{\mu_\eta\over 9}  \right )  \Biggr\}\, . 
\ea 
 
\noindent 
At the one--loop level, the 27--plet $I=0$ function 
$\tilde{a}_0^{(27)}$ is 
 
\ba 
\label{ONELOOP_27} 
&& \hspace{-0.8cm}\tilde{a}_0^{(27)}=  
 -{G_F\over \sqrt{2}} V^{\phantom{\ast}}_{ud}V^\ast_{us}  
{\sqrt{2}} f_\pi   {1\over 9}\, g_{27} \; \Biggl\{  1 
- {1\over 2} \left( 2 s - M_\pi^2 \right) B(M_\pi^2,M_\pi^2,s) 
\Biggr. \nonumber\\ 
&&\hspace{-0.8cm} \mbox{} 
+ {3\over 2} \left( s - M_K^2 + M_\pi^2\right) B(M_K^2,M_K^2,s) 
-{3\over 2}  M_\pi^2\, B(M_\eta^2,M_\eta^2,s) 
\nonumber\\ 
&&\hspace{-0.8cm} \mbox{} 
+ {(M_K^2-M_\pi^2)\over 12 M_\pi^2} \left[ 
  ( 3 M_K^2 - 4 M_\pi^2 )\, B(M_K^2,M_\pi^2,M_\pi^2 ) 
  - 4 M_K^2\, B(M_K^2,M_\eta^2,M_\pi^2 ) \right] 
\nonumber\\ 
&&\hspace{-0.8cm} \mbox{} 
+{67\over 12}\,\mu_\pi -{49\over 6}\,\mu_K  -{3\over 4}\,\mu_\eta 
  +{(M_K^2-M_\pi^2)\over 2 M_\pi^2} 
\left ( \mu_\pi+4\mu_\eta -5\mu_K \right ) 
+{M_\pi^2\over 2 M_K^2}\left ( \mu_\eta  - \mu_\pi\right ) 
\nonumber\\ 
&&\hspace{-0.8cm} \mbox{} 
+ {1\over f^2} \biggl[ 
-16 (2M_K^2+M_\pi^2)\, L_4^r(\mu)  -4 (M_K^2+3M_\pi^2)\, L_5^r(\mu) 
 +  M_K^2\, C^{27}_1(\mu) 
\biggr.\nonumber\\ && \mbox{} 
\Biggl.\biggl. 
 +M_\pi^2\, C^{27}_4(\mu) 
- {M_\pi^4\over M_K^2}\, C^{27}_6(\mu)  + s\, C^{27}_5(\mu) 
\biggr]\Biggr\}\, , 
\ea 
 
\noindent and 
 
\ba 
\label{ONELOOP_27_D} 
&& \hspace{-0.8cm}\delta\tilde{a}_0^{(27)}=  
 -{G_F\over \sqrt{2}} V^{\phantom{\ast}}_{ud}V^\ast_{us}  
{\sqrt{2}} f_\pi   {1\over 9}\, g_{27}\; \Biggl\{ 
{1\over 2} M_\pi^2\left[ 4 B(M_\eta^2,M_\eta^2,s) - 3 B(M_K^2,M_K^2,s) \right] 
 \Biggr.   \nonumber\\ 
&&\hspace{-0.8cm} \mbox{} 
-{1\over 12} (5 M_K^2+ 4 M_\pi^2)\, B(M_K^2,M_\pi^2,M_\pi^2) 
-{1\over 3} M_K^2\, B(M_K^2,M_\eta^2,M_\pi^2) 
 \nonumber\\ 
&&\hspace{-0.8cm} \mbox{} 
+ {1\over f^2 }\left[M_\pi^2\, C^{27}_2(\mu) 
  + M_K^2\, C^{27}_3(\mu)   
  +{M_\pi^4\over M_K^2}\, C^{27}_6(\mu) \right] 
    -{5\over 6}\,\mu_\pi-2\,\mu_\eta +{13\over 6}\,\mu_K 
\nonumber\\ 
&& \hspace{-0.8cm} \mbox{} 
\Biggl. -{M_\pi^2\over M_K^2 -M_\pi^2} 
\left ( {\mu_\pi\over 2} +{3\over 2}\mu_\eta -2\mu_K 
 \right) 
     +{M_\pi^4\over 2 M_K^2(M_K^2 -M_\pi^2)}\left( \mu_\eta - \mu_\pi 
\right )  \Biggr\}\, . 
\ea 

\noindent Finally, the  $I=2$ $\tilde{a}_2$ function is 
 
\ba 
\label{ONELOOP_2} 
&& \hspace{-0.8cm}\tilde{a}_{2}= 
-{G_F\over \sqrt{2}} V_{ud}^{\phantom{\ast}}V^\ast_{us}\, 
 {10\over 9}f_\pi\, g_{27} \;\Biggl\{ 1 
+ {1\over 2} \left(  s - 2 M_\pi^2 \right) B(M_\pi^2,M_\pi^2,s) 
\Biggr. \nonumber\\ 
&&\hspace{-0.8cm} \mbox{} 
+  {(M_K^2-M_\pi^2)\over 24 M_\pi^2} \left[ 
  ( 15 M_K^2 - 8 M_\pi^2 )\, B(M_K^2,M_\pi^2,M_\pi^2 ) 
  + M_K^2\, B(M_K^2,M_\eta^2,M_\pi^2 ) \right] 
\nonumber\\ 
&&\hspace{-0.8cm} \mbox{} 
 -{23\over 12}\,\mu_\pi-{1\over 4}\,\mu_\eta-{7\over 6}\, 
\mu_K +{(M_K^2-M_\pi^2)\over 4 M_\pi^2} 
\left ( 5\mu_\pi-\mu_\eta -4\mu_K  \right ) 
\nonumber\\ 
&&\hspace{-0.8cm} \mbox{} + 
{1\over f^2} \biggl[ 
 -16 (2M_K^2+M_\pi^2) L_4^r(\mu)  -4 (M_K^2+3M_\pi^2) L_5^r(\mu) 
+ M_K^2\,\bar{C}^{27}_1(\mu) 
\biggr.\nonumber\\ 
&&  \Biggl.\biggl.\mbox{} 
+ M_\pi^2\,\bar{C}^{27}_4(\mu)  
+  s\,\bar{C}^{27}_5(\mu) 
 \biggr]\Biggr\}\, , 
\ea 
 
\noindent while $\delta\tilde{a}_{2}$ is given by 
 
\ba 
\label{ONELOOP_2_D} 
&& \hspace{-1.6cm}\delta\tilde{a}_{2}= 
-{G_F\over \sqrt{2}} V_{ud}^{\phantom{\ast}}V^\ast_{us}\, 
 {10\over 9} f_\pi\, g_{27}\; \Biggl\{ 
 {1\over f^2} \left[ M_\pi^2\, \bar{C}^{27}_2(\mu) 
 + M_K^2\, \bar{C}^{27}_3(\mu) \right] 
\Biggr. \nonumber\\ 
&&\hspace{-0.8cm} \Biggl. \mbox{} 
+ {1\over 24} \biggl[ M_K^2\, B(M_K^2,M_\eta^2,M_\pi^2) 
 - (M_K^2 + 8 M_\pi^2)\, B(M_K^2,M_\pi^2,M_\pi^2)\biggr] 
\nonumber\\ 
&&\hspace{-0.8cm} \Biggl. \mbox{} 
 - {1\over 12} (\mu_\pi  + 3\mu_\eta + 4 \mu_K ) 
 +{M_\pi^2\over M_K^2-M_\pi^2} 
\left ( 2\mu_\pi -\mu_\eta -\mu_K \right ) \Biggr\}\, . 
\ea 
 
The one-loop function $B(M_1^2,M_2^2,p^2)$ is defined as follows 
 
\ba 
\lefteqn{B(M_1^2,M_2^2,p^2)  = 
\frac{1}{16 \pi^2 f^2} \left[ -1 + \ln \left( \frac{M_2^2}{\mu^2} 
\right) + \frac{1}{2} \ln \left( \frac{M_1^2}{M_2^2}\right) 
\left( 1+ \frac{M_1^2}{p^2} - \frac{M_2^2}{p^2} \right)\right. } && 
\nonumber \\ 
&& \mbox{} + \left. {1\over 2} 
\lambda^{1/2}\left( 1, \frac{M_1^2}{p^2}, \frac{M_2^2}{p^2} \right) 
\ln \left[ \frac{p^2 - M_1^2 - M_2^2 
+ \lambda^{1/2}\left( p^2, M_1^2, M_2^2 \right)} 
{p^2 -M_1^2 - M_2^2 -\lambda^{1/2}\left( p^2, M_1^2, M_2^2 \right)} 
\right] \right] \, ,  \qquad\quad 
\nonumber \\ 
\ea 
 
\noindent for $p^2 > (M_1+M_2)^2$  and $p^2 \le (M_1-M_2)^2$,  while 
 
\ba 
\lefteqn{B(M_1^2,M_2^2,p^2) = 
\frac{1}{16 \pi^2 f^2} \left[ -1 + \ln \left( \frac{M_2^2}{\mu^2} 
\right) + \frac{1}{2} \ln \left( \frac{M_1^2}{M_2^2}\right) 
\left( 1+ \frac{M_1^2}{p^2} - \frac{M_2^2}{p^2} \right) 
\right. } &&\nonumber \\ 
&&\mbox{} - \left. {1\over 2} 
\sqrt{-\lambda\left( 1, \frac{M_1^2}{p^2}, \frac{M_2^2}{p^2} \right)} 
\arctan \left[ \frac{(p^2-M_1^2-M_2^2) 
\sqrt{- \lambda(p^2,M_1^2,M_2^2)}}{(p^2-M_1^2-M_2^2)^2-2 M_1^2 M_2^2} 
\right] \right] 
\nonumber \\ 
\ea 
 
\noindent for $(M_1-M_2)^2  <  p^2 \le (M_1+M_2)^2$ and 
$-\pi/2 < \arctan \, (x) < \pi/2$. The function $\lambda(x,y,z)$ is given by  
\ba 
\lambda(x,y,z) &=& (x+y-z)^2 -4 xy \, . 
\ea 
The function $B(M_1^2,M_2^2,p^2)$ is related to the function 
$\bar{J}(s)$ introduced in ref.~\cite{GL:85} in the following way 
\be 
f^2\, B(M_1^2,M_2^2,p^2)= -\bar{J}(s) + {1\over 16\pi^2}\left ( 
 \ln  \frac{M_2^2}{\mu^2} + {M_1^2\over M_1^2-M_2^2} 
\ln {M_1^2\over M_2^2} \right )\, . 
\ee 
For $M_1=M_2\equiv m$ one gets 
\be 
f^2\, B(M^2,M^2,p^2)= -\bar{J}(s) + {1\over 16\pi^2}\left ( 
 \ln  \frac{M^2}{\mu^2} + 1 \right )\, , 
\ee 
where $\bar{J}(s)$ for $M_1=M_2$ has been given in eq.~\eqn{eq:Jb_PP}. 
 
The parameters $\mu_P$ \ ($ P=\pi ,\, K,\, \eta$) 
contain a logarithmic dependence on 
the chiral renormalization scale $\mu$ generated by one-loop corrections. 
They are defined as 
\be 
\mu_P = {M_P^2\over 32\pi^2 f^2} \;\ln {M_P^2\over \mu^2}\, . 
\ee 
The explicit $\mu$--dependence of $\mu_P$ and the  
functions $B(M_1^2,M_2^2,p^2)$ is canceled by the local 
contributions $L_i^r$ ($i= 4,\, 5$), $C^8_i$ ($i=1,\ldots 6$), 
$C^{27}_i$ ($i=1,\ldots 6$) and $\bar{C}^{27}_i$ ($i=1,\ldots 5$).

\section{Comments on recent literature} 
\setcounter{equation}{0} 
 
The ideas put forward in our first letter \cite{SHORT} have been 
further discussed in several recent papers by other authors. We 
would like to make here some brief comments on these works. 
 
\subsection{} 
 
It has been pointed out in ref.~\cite{BCFIMS:00}  
that the Omn\`es exponential depends on the chosen subtraction point.  
In that reference the FSI enhancement is minimized,  
by taking the highest subtraction point below the physical cut,  
$s_0=4 M_\pi^2$.
This trivial fact is then used to argue that our evaluation of FSI 
corrections is unreliable. Our detailed analysis of the subtraction 
point dependence in section~\ref{Scalar} shows that this 
claim is unfounded. Taking a higher value of $s_0$ one 
is just shifting FSI corrections from the Omn\`es exponential 
to the amplitude in front, but the physical result is of course the same. 
At $s_0=4 M_\pi^2$ there is a large one-loop correction to the amplitude, 
which has been overlooked in ref.~\cite{BCFIMS:00}.  
Moreover, $s_0=4 M_\pi^2$ is a bad 
choice of subtraction point, because the corresponding 
Taylor expansion has zero convergence radius \cite{TR:00}.

Ref.~\cite{BCFIMS:00} states that it is not precisely known at which value
of $s$ existing lattice estimates correspond to. It is suggested that
future lattice calculations could obtain the $K\to\pi\pi$ weak matrix elements
at threshold ($s_0=4 M_\pi^2$) and the $\Re_I(M_K^2,s_0)$ correction factors
could then be used to get the physical amplitudes.

The $s$ ambiguity mentioned in ref.~\cite{BCFIMS:00} is not present in the
low-energy chiral expansion.
Our $\varepsilon'/\varepsilon$ calculation is based on a
large--$N_C$ evaluation of the couplings of the $\Delta S=1$ $\chi$PT
Lagrangian. Once these chiral couplings are determined the
$K\to\pi\pi$ amplitudes can in principle be computed at any value of $s$.
Higher--order chiral corrections are of course smaller at lower values 
of momenta,
which makes advisable to use the chiral expansion al low $s$ values.

\subsection{} 
 
At lowest order in $\chi$PT, the four--quark operator $Q_8$ induces the 
$O(p^0)$ chiral term proportional to the coupling $g_{EW}$. 
The corresponding one--loop correction has been analyzed in ref.~\cite{CG:00}, 
where a small positive contribution is obtained. 
This result agrees with a recent dispersive calculation of $B_8^{(3/2)}$ 
\cite{DG:00}, which finds 
$B_8^{(3/2)} = 1.11\pm 0.16\pm 0.23$. 
 
FSI generate instead a small suppression of the $I=2$ 
amplitude. However, there are other chiral corrections not related to 
FSI which appear at the one-loop level; they are included in the 
value of $a_2(0)$. Since for $I=2$ the FSI effect is small, other 
correction could be equally important and even reverse the sign 
of the correction for the physical amplitude $a_2(M_K^2)$. 
A detailed one-loop analysis will be presented elsewhere 
\cite{EIMNP:00}. 

\subsection{} 
 
A simplified (the dispersive integral over the phase-shift 
is not exponentiated) version of our $\Re_I$ factors 
has been used in ref.~\cite{PA:99}, which advocates a different and 
conceptually incorrect 
interpretation of the chiral corrections related to FSI. 
 
In this reference a non-subtracted dispersion relation is used. The 
resulting divergence in $I^I_{0,0}(s)$ 
is regulated cutting the dispersive integral at the upper edge 
$\bar z$, and making the ad-hoc identification 
$\bar z = \nu^2$, with $\nu$ the short--distance scale governing 
the Wilson coefficients of the effective four--quark Hamiltonian 
(\ref{eq:Leff}). This generates a $\nu$ dependence in the dispersive integral 
which is claimed to cancel the renormalization--scale dependence of the Wilson 
coefficients.
 
The choice $\bar z = \nu^2$ and the associated identification 
of infrared and ultraviolet logarithms is arbitrary 
and cannot be correct. The FSI logarithms have 
nothing to do with the underlying short--distance physics. The 
Omn\`es factor $\Omega_I(s,s_0)$ relates the isospin amplitudes 
at two different points $s_0$ and $s$, but is unable to fix the 
global normalization. The short--distance information is hidden 
in $a_I(0)$ which, moreover, is independent of the scale $\nu$. 
Thus, the cancellation of $\nu$ dependences must be 
accomplished even in the absence of FSI. 
 
The argument can be better seen analyzing the scalar form factor, 
which has the same FSI logarithms but a different short--distance 
contribution. In fact, instead of working with hadronic matrix 
elements of the scalar current, we can take the corresponding 
matrix elements of the divergence of the associated vector current. 
Both quantities are trivially related by a quark--mass 
factor, through a Ward identity. The FSI phenomena and the 
associated $\chi$PT logarithms are of course identical; but now, 
there is no anomalous dimension. Since there is no 
short--distance renormalization scale $\nu$, the identification 
$\bar z = \nu^2$ is then meaningless. 
 
The same argumentation can be repeated with the Omn\`es summation 
of FSI effects in the pion vector form factor \cite{PLBFF}. Again, 
this is a renormalization--group invariant quantity (no anomalous 
dimension) and does not make any sense to identify FSI phenomena 
with non-existing short--distance logarithms.

 
\end{document}